\documentstyle[draft,psfig]{mn} 

\newif\ifAMStwofonts

\ifoldfss
  \ifCUPmtlplainloaded \else
    \NewTextAlphabet{textbfit} {cmbxti10} {}
    \NewTextAlphabet{textbfss} {cmssbx10} {}
    \NewMathAlphabet{mathbfit} {cmbxti10} {} 
    \NewMathAlphabet{mathbfss} {cmssbx10} {} 
  \fi
  \ifAMStwofonts
    \ifCUPmtlplainloaded \else
      \NewSymbolFont{upmath} {eurm10}
      \NewSymbolFont{AMSa} {msam10}
      \NewMathSymbol{\upi}     {0}{upmath}{19}
      \NewMathSymbol{\umu}     {0}{upmath}{16}
      \NewMathSymbol{\upartial}{0}{upmath}{40}
      \NewMathSymbol{\leqslant}{3}{AMSa}{36}
      \NewMathSymbol{\geqslant}{3}{AMSa}{3E}

      \let\leq=\leqslant 
      \let\geq=\geqslant 
    \fi
  \fi
\fi 

\ifnfssone
  \newmathalphabet{\mathit}
  \addtoversion{normal}{\mathit}{cmr}{m}{it}
  \addtoversion{bold}{\mathit}{cmr}{bx}{it}
  \newmathalphabet{\mathbfit} 
  \addtoversion{normal}{\mathbfit}{cmr}{bx}{it}
  \addtoversion{bold}{\mathbfit}{cmr}{bx}{it}
  \newmathalphabet{\mathbfss} 
  \addtoversion{normal}{\mathbfss}{cmss}{bx}{n}
  \addtoversion{bold}{\mathbfss}{cmss}{bx}{n}
  \ifAMStwofonts
    \ifCUPmtlplainloaded \else
      \UseAMStwoboldmath
      \makeatletter
      \new@mathgroup\upmath@group
      \define@mathgroup\mv@normal\upmath@group{eur}{m}{n}
      \define@mathgroup\mv@bold\upmath@group{eur}{b}{n}
      \edef\UPM{\hexnumber\upmath@group}
      \new@mathgroup\amsa@group
      \define@mathgroup\mv@normal\amsa@group{msa}{m}{n}
      \define@mathgroup\mv@bold\amsa@group{msa}{m}{n}
      \edef\AMSa{\hexnumber\amsa@group}
      \makeatother
      \mathchardef\upi="0\UPM19
      \mathchardef\umu="0\UPM16
      \mathchardef\upartial="0\UPM40
      \mathchardef\leqslant="3\AMSa36
      \mathchardef\geqslant="3\AMSa3E

      \let\leq=\leqslant 
      \let\geq=\geqslant 
    \fi
  \fi
\fi 

\ifnfsstwo
  \DeclareMathAlphabet{\mathbfit}{OT1}{cmr}{bx}{it}
  \SetMathAlphabet\mathbfit{bold}{OT1}{cmr}{bx}{it}
  \DeclareMathAlphabet{\mathbfss}{OT1}{cmss}{bx}{n}
  \SetMathAlphabet\mathbfss{bold}{OT1}{cmss}{bx}{n}
  \ifAMStwofonts
    \ifCUPmtlplainloaded \else
      \DeclareSymbolFont{UPM}{U}{eur}{m}{n}
      \SetSymbolFont{UPM}{bold}{U}{eur}{b}{n}
      \DeclareSymbolFont{AMSa}{U}{msa}{m}{n}
      \DeclareMathSymbol{\upi}{0}{UPM}{"19}
      \DeclareMathSymbol{\umu}{0}{UPM}{"16}
      \DeclareMathSymbol{\upartial}{0}{UPM}{"40}
      \DeclareMathSymbol{\leqslant}{3}{AMSa}{"36}
      \DeclareMathSymbol{\geqslant}{3}{AMSa}{"3E}

      \let\leq=\leqslant 
      \let\geq=\geqslant 
    \fi
  \fi
\fi 

\ifCUPmtlplainloaded \else
  \ifAMStwofonts \else 
    \def\upi{\pi}
    \def\umu{\mu}
    \def\upartial{\partial}
  \fi
\fi

\title[Black hole--neutron star coalescence]{Newtonian hydrodynamics
of the coalescence of black holes with neutron stars -- IV: Irrotational
binaries with a soft equation of state}

\author[W.~H. Lee]
{William H. Lee \\
Instituto de Astronom\'{\i}a, Universidad Nacional Aut\'{o}noma
de M\'{e}xico, Apdo. Postal 70--264, Cd. Universitaria, 04510
M\'{e}xico D.F.\\
}

\pagerange{\pageref{firstpage}--\pageref{lastpage}}
\pubyear{2001}

\begin{document}

\maketitle

\label{firstpage}


\begin{abstract}
   We present the results of three--dimensional hydrodynamical
   simulations of the final stages of inspiral in a black
   hole--neutron star binary, when the separation is comparable to the
   stellar radius. We use a Newtonian Smooth Particle Hydrodynamics
   (SPH) code to model the evolution of the system, and take the
   neutron star to be a polytrope with a soft (adiabatic index
   $\Gamma=2$ and $\Gamma=5/3$) equation of state and the black hole
   to be a Newtonian point mass. The only non--Newtonian effect we
   include is a gravitational radiation back reaction force, computed
   in the quadrupole approximation for point masses. We use
   irrotational binaries as initial conditions for our dynamical
   simulations, which are begun when the system is on the verge of
   initiating mass transfer and followed for approximately 23~ms. For
   all the cases studied we find that the star is disrupted on a
   dynamical time--scale, and forms a massive ($M_{disc}\approx
   0.2$~$M_{\odot}$) accretion torus around the spinning (Kerr) black
   hole. The rotation axis is clear of baryons (less than
   $10^{-5}M_{\odot}$ within $10^{\circ}$) to an extent that would not
   preclude the formation of a relativistic fireball capable of
   powering a cosmological gamma ray burst. Some mass (the specific
   amount is sensitive to the stiffness of the equation of state) may
   be dynamically ejected from the system during the coalescence and
   could undergo r--process nucleosynthesis. We calculate the
   waveforms, luminosities and energy spectra of the gravitational
   radiation signal and show how they reflect the global outcome of
   the coalescence process.
\end{abstract}
   


\section{Introduction and motivation}

In binary systems, the emission of gravitational waves and
accompanying loss of angular momentum will lead to a decrease in the
orbital separation, and coalescence will occur if the decay time is
small enough (less than the Hubble time). For binaries made of neutron
stars, PSR~1913+16 being the most famous example, this consequence of
general relativity has been observed indirectly~\cite{hulse75} (see
also Wolszczan~\shortcite{wolszczan91} for the case of PSR~1534+12),
and the change in orbital period matches the theoretical predicion to
very high accuracy~\cite{taylor,stairs98}. Given their present--day
orbital periods (on the order of 10 hours), these systems will
eventually merge. The final stages of the coalescence present an
opportunity to study the equation of state at very high densities (the
system is in effect a giant accelerator), and will undoubtedly produce
a strong electromagnetic and gravitational wave signal containing some
of this information. No black hole--neutron star binary systems are
known yet, but population synthesis
studies~\cite{LS76,narayan91,tutukov,lipunov,bethe,portyun,belczynski,kalogera01a,kalogera01b}
over the past 25 years lead one to believe that their rate is
comparable to that of double neutron star binaries, and is on the
order of $10^{-6}-10^{-5}$ per year per galaxy.

Solving this problem completely clearly requires detailed hydrodynamic
modeling in three dimensions, radiation transport, a realistic
equation of state, and general relativity. As such, it must be
approached in stages, with successive approximations depending on the
aspect of the general problem one wishes, and is able, to solve.

Compact binaries are expected to be sources of gravitational radiation
observable by detectors such as LIGO \cite{abram92} and VIRGO
\cite{brad} as the inspiral occurs. The signal can be approximated as
that of point--masses and calculated accurately using post--Newtonian
expansions when the separation is large, compared with the stellar
radius \cite{kidder92,cutler,blanchet95}. When the distance between
the stars becomes comparable to their radii, hydrodynamical modeling
becomes essential. The theoretical study of the tidal disruption of a
neutron star by a black hole was addressed many years ago
\cite{wheeler71,LS74,LS76}, and numerical hydrodynamical studies of
binary neutron star coalescence were begun somewhat more recently,
initially by Oohara \& Nakamura \shortcite{oohara89,oohara90,oohara92}
and Nakamura \& Oohara \shortcite{nakamura89,nakamura91}. The work of
Chandrasekhar \shortcite{ch69} on incompressible ellipsoids was
generalized to the compressible case in the Newtonian regime by Lai,
Rasio \& Shapiro \shortcite{LRSb}, using a polytropic equation of
state, who showed that tidal effects alone could produce a
de--stabilization of the orbit in certain situations
\cite{LRSa}. Rasio \& Shapiro \shortcite{RS92,RS94,RS95} then
performed a series of dynamical simulations to study the coalescence
of two neutron stars, using Smooth Particle Hydrodynamics (SPH), while
Zhuge, Centrella \& McMillan \shortcite{zhuge94,zhuge96} focused on
the gravitational waves spectrum. Both of these groups used a
polytropic equation of state throughout their analysis. The
thermodynamical details of the process were studied by Davies et
al. \shortcite{davies}, Ruffert, Janka \& Sch\"{a}{fer
\shortcite{ruffert96}, Ruffert et al. \shortcite{ruffert97} and
Rosswog et al. \shortcite{rosswog99,rosswog00}, by using the equation
of state of Lattimer \& Swesty \shortcite{LS}. This work was all done
using a Newtonian or modified Newtonian approach (by including
gravitational radiation reaction in different ways in the
calculations), and we note that the thermodynamic details are of
little importance for the emission of gravitational waves, since it is
concernced primarily with the motion of bulk matter at high
densities. More recently, there have been advances in making
post--Newtonian calculations of initial conditions \cite{lombardi97}
and mergers \cite{faber00,faber01,ayal01}, and also in including
general relativity
\cite{wilson96,baumgarte97,oohara99,shibata99,shibata00,uryu00,usui00,gourgoulhon01}.

\begin{figure}
\psfig{width=7.5cm,file=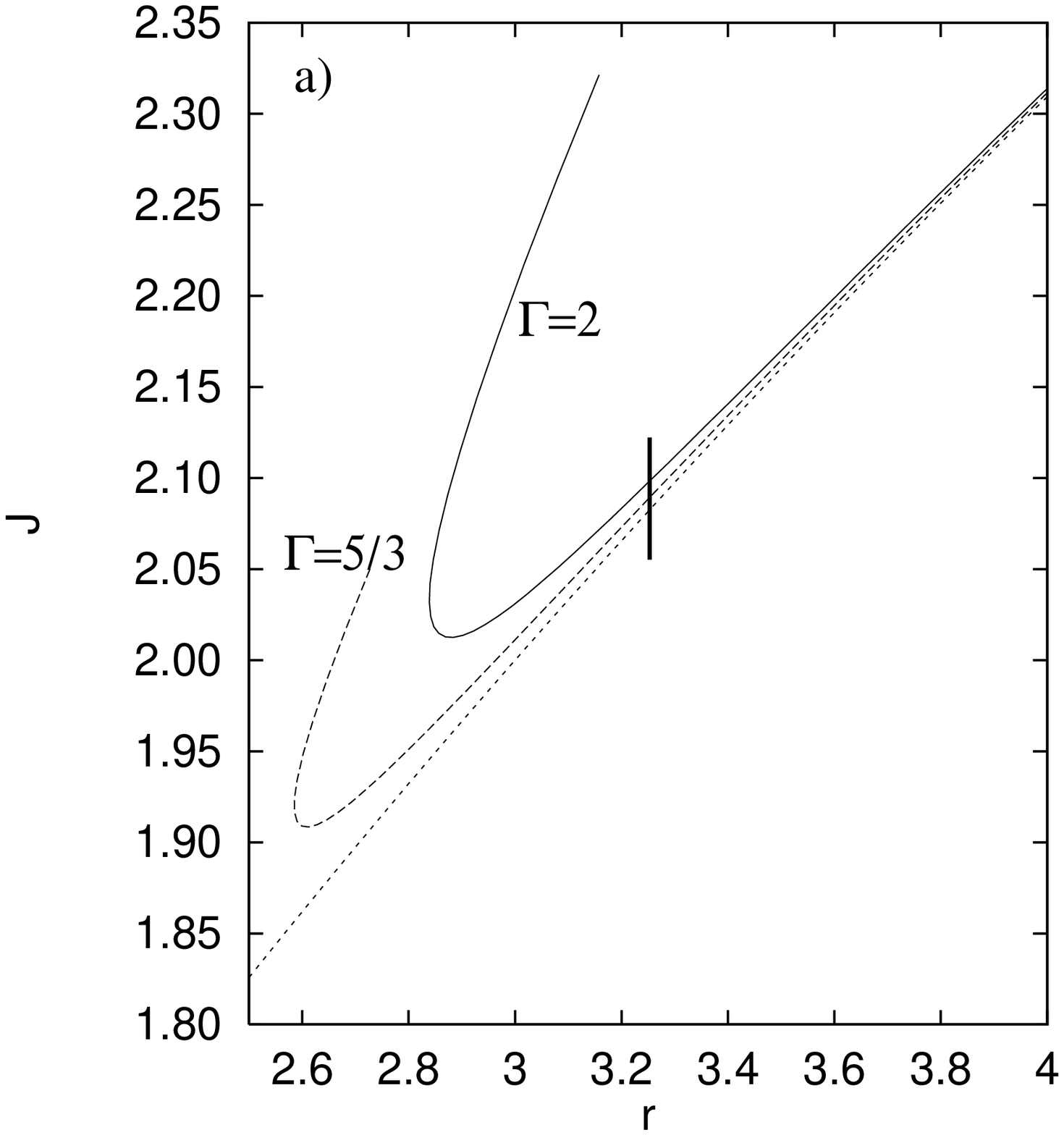,angle=0,clip=}
\end{figure}

\begin{figure}
\psfig{width=7.5cm,file=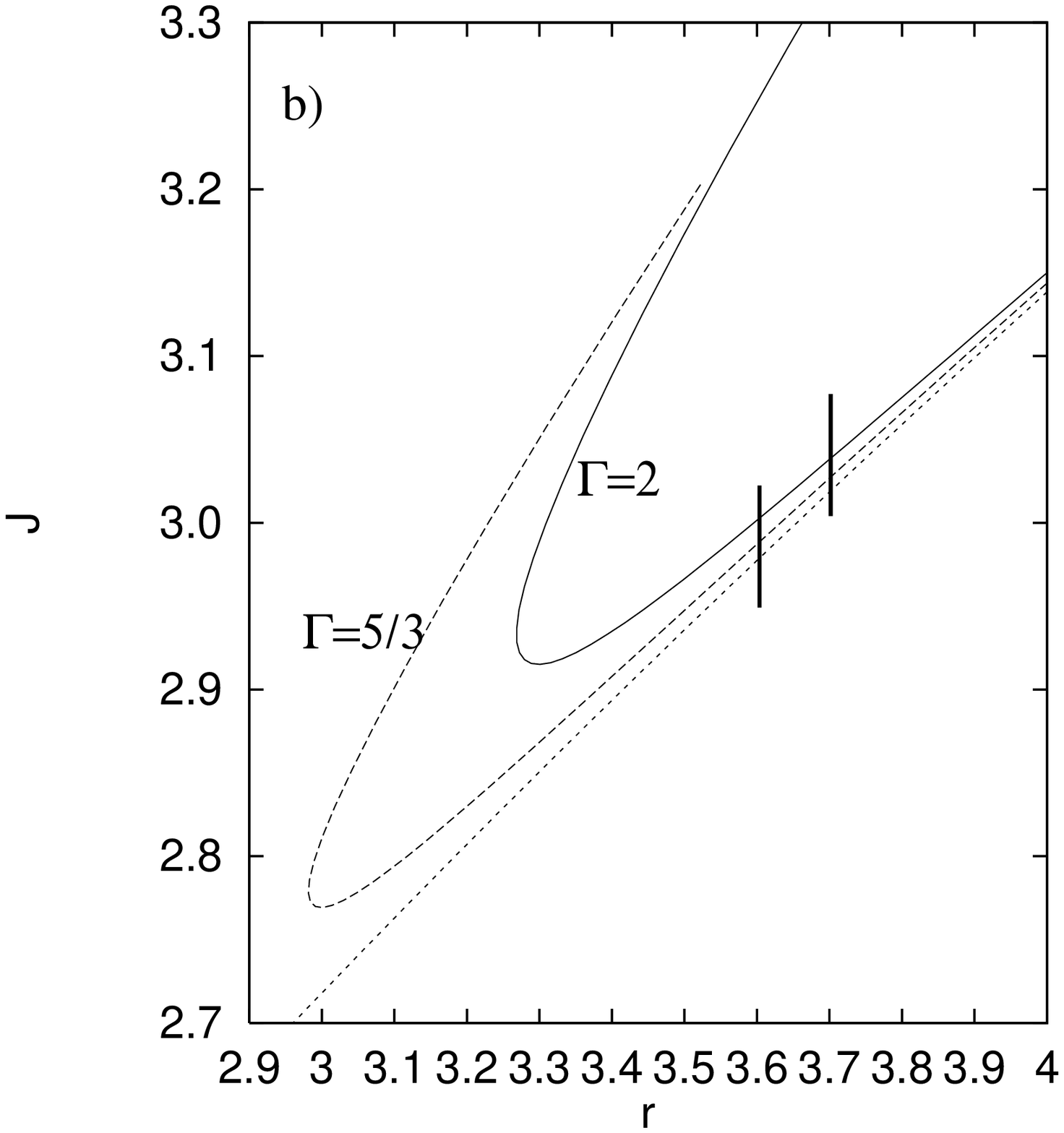,angle=0,clip=}
\end{figure}
\setcounter{figure}{0}
\begin{figure}
\psfig{width=7.5cm,file=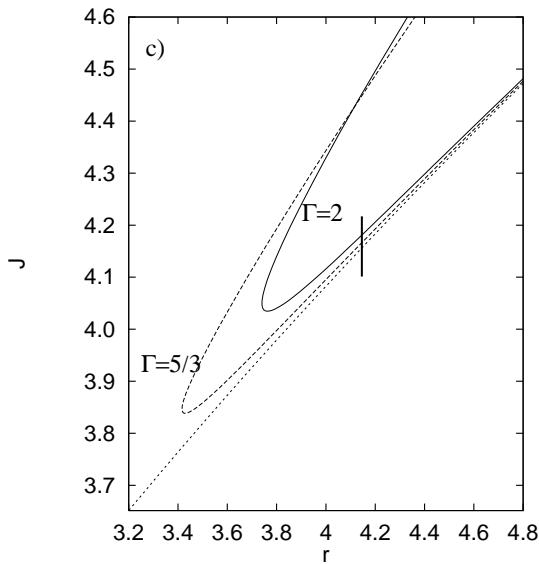,angle=0,clip=}
\caption{Total angular momentum as a function of binary separation for
irrotational Roche--Riemann binaries using the method of LRSb for
$\Gamma=2$ (solid lines) and $\Gamma=5/3$ (dashed lines) for (a)
q=0.5; (b) q=0.31; (c) q=0.2. The dotted lines correspond to the
solution computed for point masses in Keplerian orbits. The thick
vertical lines mark the initial separations used for the dynamical
runs.}
\label{jr}
\end{figure}

The gamma ray burts (GRBs) are now believed to be at cosmological
distances \cite{meegan}, after the discovery of optical afterglows
\cite{rees97a} in the last few years that have established their
redshifts \cite{metzger97,djorgovski98,kulkarni98,kulkarni99}. Reviews
have been given by Fishman \& Meegan \shortcite{fishman95} and van
Paradijs, Kouveliotou \& Wijers \shortcite{vanpar00}. Observations
have shown that (i) there is a bimodality in burst durations
\cite{kouveliotou}, with classes of short ($t_{burst}\simeq 0.5$~s)
and long ($t_{burst}\simeq 40$~s) bursts; (ii) the energy release if
one assumes isotropy is on the order of $10^{51}-10^{53}$~erg; (iii)
many bursts show variability on very short (ms) time--scales; (iv) at
least some bursts are beamed, implying a lower energy release than
isotropic emission would lead one to believe (see e.g. Harrison et
al.~\shortcite{harrison99}; Stanek et al.~\shortcite{stanek99}; Frail
et al.~\shortcite{frail01}). The preferred model for GRBs involves the
expansion of a relativistic fireball \cite{rm92,rees93} which would
produce the gamma rays through relativistic shocks and subsequent
synchrotron radiation. The fireball would presumably originate from a
central engine capable of accomodating the observational requirements
mentioned above. A variety of progenitors involving compact objects
have been suggested, see e.g. Fryer, Woosley \& Hartmann
\shortcite{fryer99a}. Many of them invoke an accretion torus around a
black hole, originating from a double neutron star coalescence
\cite{goodman86,paczynski86,eichler,narayan92}, where the central
object would presumably collapse to a black hole, the merger of a
neutron star, white dwarf or Helium core with a black hole
\cite{paczynski91,fryer99b,zhang01}, or a ``failed supernova'' or
collapsar \cite{woosley93,macfadyen}, where a massive star collapses
but produces a black hole instead of a neutron star at its center.
These systems would produce a GRB by tapping the binding energy of the
disc through neutrino emission
\cite{goodman87,jaro93,jaro96,witt,thompson,moch93,moch95,popham,ruffert99},
or the rotational energy of the black hole, through the Blandford \&
Znajek~\shortcite{blandford77} mechanism, producing so--called
Poynting jets \cite{rees97b,mesz99,lee00}. Another class of models
also involves neutron stars, but would power the GRB through the
catastrophic release of rotational energy via intense magnetic fields
\cite{usov92,kr98,spruit99,rtk00}, or even through intense neutrino
emission in a neutron star binary before the coalescence, because of
tidal heating and compression \cite{salmonson01}.

The ejection of neutron star matter to the interstellar medium during
a dynamical coalescence might contribute to the abundances of heavy
elements \cite{LS74,LS76,symb82,eichler}, in addition to the amounts
expected from supernova explosions \cite{meyer97,frei99a}. This
question has been addressed in the numerical calculations of double
neutron star mergers of Rosswog et al. \shortcite{rosswog99,rosswog00}
and by Freiburghaus et al. \shortcite{frei99b}. If the rates of black
hole--neutron star mergers are comparable, it is possible that these
systems might also contribute in the same way to the galactic
abundances.

Our work on merging black hole--neutron star binaries began with
low--resolution simulations \cite{acta} that used a stiff polytropic
equation of state. The results initially led us to believe that, if
proved true, these models were not likely to produce cosmological
gamma ray bursts, because of excessive baryon contamination. We
gradually increased our numerical resolution, using essentially
Newtonian physics (except for our treament of gravitational radiation
reaction, see below, \S~\ref{method}), and treated tidally locked
binaries with stiff and soft equations of state \cite{LKI,LKII}, and
irrotational binaries with a stiff equation of state \cite{LKIII},
always using a polytrope to model the initial neutron star. It became
apparent early on that our initial suspicions were unfounded, and that
indeed these systems were promising candidates for the central engines
of GRBs \cite{hunt4,kl98}. We also found that for very stiff equations
of state (see papers I and III) the neutron star could avoid immediate
tidal disruption, and that this would be reflected in the
gravitational wave signal. Additionally, a substantial amount of mass
could be ejected to the interstellar medium, and potentially undergo
r--process nucleosynthesis, thus contributing to the abundances of
heavy elements. Recently, Janka et al. \shortcite{janka99} used the
same formalism that Ruffert et al. \shortcite{ruffert96,ruffert97} had
employed for binary neutron star coalescence studies to simulate the
merger of a black hole with a neutron star. Their calculations have
revealed the same qualitative aspects of the process which we have
found, with differences due mainly to the different formalism used for
gravitational radiation reaction and their use of a different equation
of state \cite{LS}.

This paper is the last in the series that
has used the approach briefly described above (and detailed below in
\S~\ref{method}), having mapped the parameter space we intended to
explore by varying the stiffness of the equation of state, the initial
mass ratio in the binary and the distribution of angular momentum in
the system (using tidally locked and irrotational binaries as initial
conditions). A short exposition on the numerical method and initial
conditions is given in \S~\ref{method} and \S~\ref{initial} (for
details concerning the implementation we refer the reader to the
longer corresponding sections in paper~III and the appendix in
paper~I), followed by our results in \S~\ref{results}. The effect
different choices of initial conditions can have on the dynamical
coalescence is presented in \S~\ref{sphere}, and a summary and
discussion is given in \S~\ref{discussion}.

\section{Numerical method} \label{method}

For the calculations presented in this work, we have used the method
known as Smooth Particle Hydrodynamics (SPH) (see Monaghan 1992 for a
review and Lee 1998 for a description of our own code). The code is
the same one that was used for our previous simulations of
irrotational black hole--neutron star binaries (paper~III).  Here we
will not discuss the code in detail, but limit the presentation to a
few basic points.

As before, the black hole is modeled as a Newtonian point mass of mass
$M_{BH}$ with an absorbing boundary at the Schwarzschild radius
$r_{Sch}=2GM_{\rm BH}/c^{2}$. Any SPH particle that crosses this
boundary is absorbed by the black hole, whose mass and momentum are
adjusted so as to ensure conservation of total mass and total linear
momentum in the system.

The neutron star is modeled as a polytrope with a soft equation of
state, so that the pressure is given by $P=K \rho^{\Gamma}$ with
$\Gamma$ and $K$ being constants (see paper~I). Unless otherwise
noted, we measure mass and distance in units of the mass and radius of
the unperturbed (spherical) neutron star (13.4~km and 1.4~$M_{\odot}$
respectively), so that the units of time, density and velocity are
\begin{eqnarray}
\tilde{t}=1.146\times 10^{-4}{\rm s}\times \left( \frac{R}{13.4~{\rm
km}}\right) ^{3/2} \left( \frac{M_{\rm NS}}{1.4~M_{\odot}}\right)
^{-1/2} \label{eq:deftunit}
\end{eqnarray}
\begin{eqnarray}
\tilde{\rho}=1.14\times 10^{18}{\rm kg~m^{-3}}\times \left(
\frac{R}{13.4~{\rm km}}\right) ^{-3} \left( \frac{M_{\rm
NS}}{1.4~M_{\odot}}\right) \label{eq:defrhounit}
\end{eqnarray}
\begin{eqnarray}
\tilde{v}=0.39 c \times \left( \frac{R}{13.4~{\rm km}}\right) ^{-1/2}
\left( \frac{M_{\rm NS}}{1.4~M_{\odot}}\right) ^{1/2} \label{eq:defvunit}
\end{eqnarray}
For the dynamical simulations presented here, we have used $N\simeq
80,000$ SPH particles to model the neutron star. The initial
(spherical) polytrope is constructed by placing the SPH particles on a
uniform three--dimensional grid with particle masses proportional to
the Lane--Emden density. This ensures that the spatial resolution is
approximately uniform throughout the fluid. This isolated star is then
allowed to relax for a period of thirty time units (as defined above)
by including a damping term linear in the velocities in the equations
of motion. The specific entropies of the particles are kept constant
during this procedure (i.e. {\em K} is constant in the equation of
state $P=K \rho^{\Gamma}$).

To perform a dynamical run, the black hole and every SPH particle are
given the velocity as determined from the corresponding initial
condition (see below) in an inertial frame, with the origin of
coordinates at the centre of mass of the system. Each SPH particle is
assigned a specific internal energy $u_{i}=K
\rho^{(\Gamma-1)}/(\Gamma-1)$ and the equation of state is changed to
that of an ideal gas, $P=(\Gamma-1)\rho u$. The specific internal
energy is then evolved according to the first law of thermodynamics,
taking into account the contribution from artificial viscosity (see
below). We vary the initial mass ratio $q=M_{\rm NS}/M_{\rm BH}$ in
the binary by adjusting the mass of the black hole only.

Artificial viscosity is used in SPH to handle the presence of shocks
and avoid particle interpenetration. As in paper~III, we use the form
of Balsara \shortcite{balsara}, which vanishes in regions of large
vorticity but retains the ability to deal with the presence of shocks
(in regions of strong compression).

We include gravitational radiation reaction in the quadrupole
approximation for point masses \cite{landau75}, with the same
implementation as described in paper~III. Namely, we apply a back
reaction force to the black hole and the self--bound core of the
neutron star, treating the latter as a point mass. The corresponding
terms in the equations of motion are switched off once the star is
tidally disrupted, when the core mass drops below 0.14~$M_{\odot}$.
We continuously compute the radiation reaction time--scale
$t_{RR}^{-1}=256G^{3}M_{\rm BH}M_{\rm core}(M_{\rm BH}+M_{\rm
core})/(5r^{4}c^{5})$ and an estimate of the current orbital period
$t_{orb}=2\pi / \sqrt{G(M_{\rm BH}+M_{\rm core})/r^{3}}$, where {\em
r} is the separation between the black hole and the centre of mass of
the core. For the typical separations and masses in the black
hole--core binary, by the time the core mass has dropped to
0.14~$M_{\odot}$, the radiation reaction time--scale is much longer
(by at least an order of magnitude) than the current orbital period.
 
\section{Initial conditions} \label{initial}

Exactly as for the results presented in paper~III, we have used
irrotational binaries for the dynamical runs shown here. This amounts
to considering the stars have no spin in an external, inertial frame
of reference. This initial condition is more realistic than that of a
tidally locked binary, because the viscosity inside neutron stars is
not large enough to maintain synchronization during the inspiral phase
\cite{bildsten,kochanek}. Essentially, the stars will coalesce with
whatever spin angular momenta they have when the binary separation is
large. Setting up accurate and self--consistent initial conditions in
this case is not a trivial matter, and we use the same approximation
as before. Namely, we apply the energy variational method of LRSb and
take the neutron star to be a compressible tri--axial Roche--Riemann
ellipsoid (see \S~8 in LRSb).

We build an initial condition by first constructing a spherical star
of given radius and mass, as described in \S~\ref{method}.  We then
use the method of LRSb to calculate the orbital angular velocity of
the binary and the semimajor axes of the Roche--Riemann ellipsoid for
the appropriate choice of adiabatic index, initial mass ratio and
binary separation (see Table~\ref{parameters}). The semimajor axes of
the fluid configuration can also be calculated from the SPH numerical
solution using
\begin{eqnarray*}
a_{i}=\sqrt{\frac{5 I_{ii}}{\kappa_{n}M_{\rm NS}}}
\end{eqnarray*}
where 
\begin{eqnarray*}
I_{ii}=\sum_{j}m_{j}(x^{i}_{j})^{2}.
\end{eqnarray*}
The stiffness of the equation of state enters these equations through
the parameter $\kappa_{n}$ ($\kappa_{n}=0.653$ for $\Gamma=2$ and
$\kappa_{n}=0.511$ for $\Gamma=5/3$). The first and second semimajor
axes of the tri--axial ellipsoid lie in the orbital plane, with the
first one along the line joining the two binary components. The third
axis is oriented perpendicular to this plane (along the axis of
rotation).  The position of each SPH particle is then re--scaled
(independently along each coordinate axis) so that the new fluid
configuration has the appropriate semimajor axes. This ellipsoid is
then used as an initial condition for the corresponding dynamical
run. The initial velocity is given by the orbital angular velocity
(for the azimuthal component) plus the radial velocity corresponding
to point--mass inspiral. The variation in total angular momentum as a
function of binary separation for irrotational Roche--Riemann binaries
(with various mass ratios and adiabatic indices in the equation of
state) is shown in Figure~\ref{jr}, as calculated using the method of
LRSb. The curves show a turning point as the separation is decreased,
indicating the presence of a dynamical instability in the system. Two
distinctions are necessary at this point. First, the ellipsoidal
approximation becomes less accurate as the separation is
decreased. This applies regardless of the value of the adiabatic
index, but is much more serious for stiff equations of state, because
the tidal effects are more pronounced. Second, the adiabatic index
does determine if the Roche limit (when mass transfer through overflow
of the lobe occurs) is reached before or after the dynamical
instability. For stiff equations of state (such as the ones shown in
papers~I \& III), the instability can be reached at or before the
Roche limit. However, for a more compressible case (see paper~II) the
inverse occurs, and it is the mass transfer process itself (which is
unstable) that is responsible for the subsequent evolution of the
system.

We have chosen the values of the initial separation for our dynamical
runs $r_{i}$ to be slightly above the turning point (see
Table~\ref{parameters}). The ellipsoidal approximation is then still
reasonable, and our equilibrium configurations have not yet reached
the point where the neutron star will overflow its Roche lobe. When
the dynamical simulation is initiated, the separation will decrease
due to the emission of gravitational waves, and mass transfer will
start promptly. The construction of full equilibrium initial
conditions at the point of Roche lobe overflow is a problem that was
addressed by Ury\={u} \& Eriguchi \shortcite{uryu99}.

The initial separations we have chosen are similar to what we have
presented before for the case of tidally locked (papers~I \& II) and
irrotational (paper~III) black hole--neutron star binaries.  We also
include in Table~\ref{parameters} the initial parameters for two runs
(C31S and D31S) that used initially spherical neutron stars for the
dynamical calculations, with a Keplerian orbital angular velocity (as
for runs A31S and B31S in paper~III). We have performed these runs to
gauge the effect non--equilibrium initial conditions will have on the
evolution of the system, and show the results in \S~\ref{sphere}.

\begin{table*}
 \caption{Basic parameters for each run. The table lists for
each run (labeled) the initial mass ratio, the adiabatic index used,
the initial orbital separation, the axis ratios for the tri--axial
ellipsoid used as an initial condition, the initial orbital angular
velocity of the binary, the time at which gravitational radiation
reaction is switched off in the simulation, the time at which the
simulation was stopped, and the initial number of particles. The runs
labeled C31S and D31S used an initially spherical neutron star
(otherwise irrotational Roche--Riemann ellipsoids were used, see text
for details).}
 \label{parameters}
 \begin{tabular}{@{}lccccccccc}
  Run & $q$ & $\Gamma$ & $r_{i}$ & $a_{2}/a_{1}$ & $a_{3}/a_{1}$  
        & $\tilde{\Omega}$  
        & $t_{rad}$
        & $t_{f}$ & $N$ \\
  C50   & 0.50 & 2.0 & 3.25 & 0.842 & 0.857 & 0.29753 & 35.12 & 200.0 
        & 81,608 \\
  C31   & 0.31 & 2.0 & 3.70 & 0.828 & 0.844 & 0.29042 & 35.80 & 200.0 
        & 81,608 \\
  C31S  & 0.31 & 2.0 & 3.70 & 1.000 & 1.000 & 0.28881 & 34.76 & 200.0 
        & 81,608 \\
  C20   & 0.20 & 2.0 & 4.15 & 0.808 & 0.826 & 0.29117 & 30.12 & 200.0 
        & 81,608 \\
  D50   & 0.50 & 5/3 & 3.25 & 0.904 & 0.911 & 0.29644 & 37.03 & 200.0 
        & 82,136 \\
  D31   & 0.31 & 5/3 & 3.60 & 0.884 & 0.892 & 0.30182 & 30.40 & 200.0 
        & 82,136 \\
  D31S  & 0.31 & 5/3 & 3.60 & 1.000 & 1.000 & 0.30094 & 29.44 & 200.0 
        & 82,136 \\
  D20   & 0.20 & 5/3 & 4.15 & 0.884 & 0.892 & 0.29037 & 26.96 & 200.0 
        & 82,136 \\

 \end{tabular}

\end{table*}

\section{Results} \label{results}

\subsection{Morphology of the mergers}

\begin{figure*}
\psfig{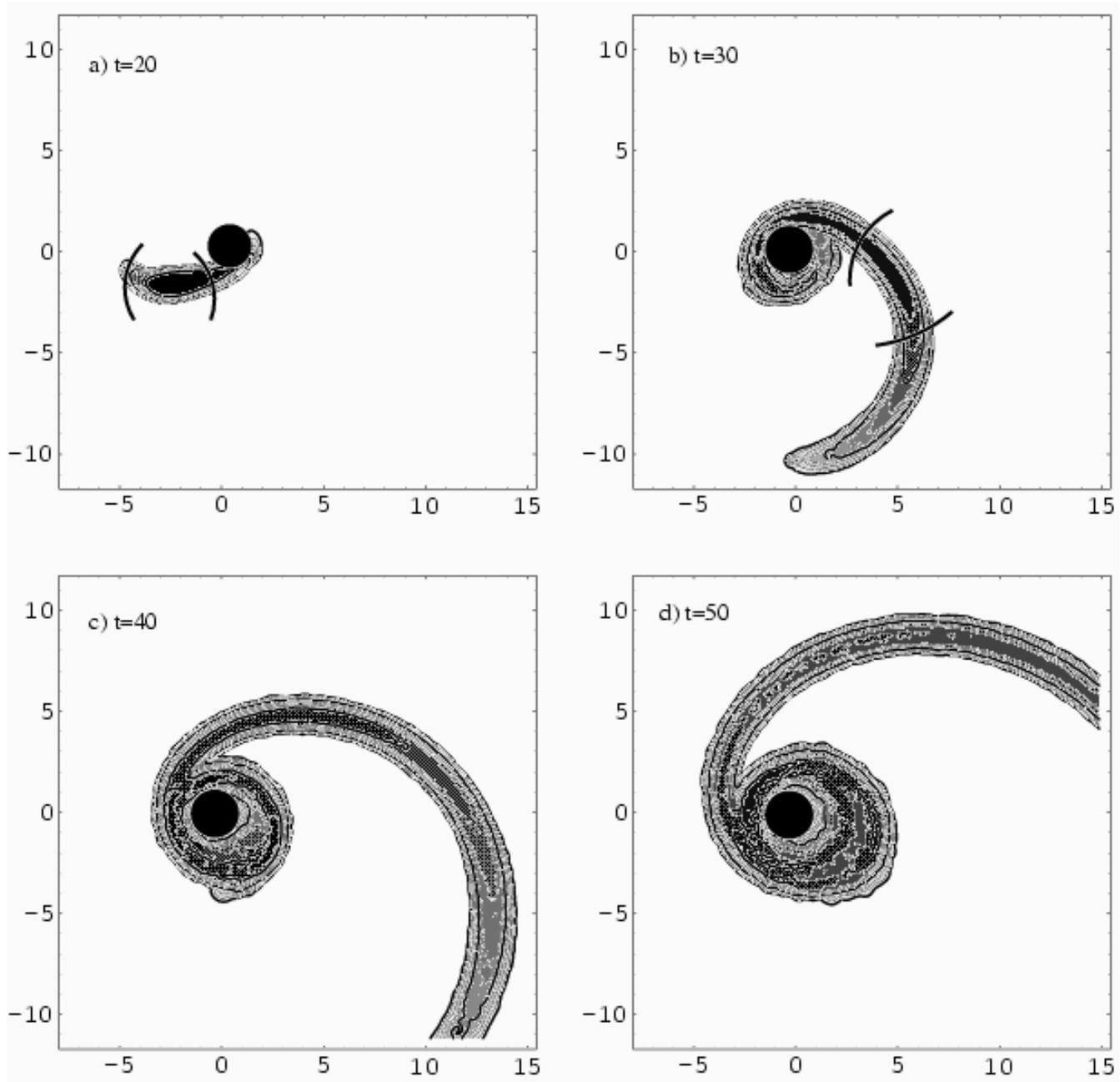}
\caption{Density contours in the orbital plane during the dynamical
simulation of the black hole--neutron star binary with initial mass
ratio $q=0.31$ and $\Gamma=2$ (run C31). The orbital rotation is
counterclockwise. All contours are logarithmic and equally spaced
every 0.25 dex. Bold contours are plotted at $\log \rho=-4,-3,-2,-1$
(if present) in the units defined in eq.~\ref{eq:defrhounit}. The
thick black arcs bound the matter that forms the core (see
\S~\ref{method}). The time for each frame is given in the units
defined in eq.~\ref{eq:deftunit}.}
\label{rhog2}
\end{figure*}

\setcounter{figure}{1}
\begin{figure*}
\psfig{width=\textwidth,file=rhoxyg2f2.eps.a,angle=0,clip=}
\caption{continued}
\end{figure*}

\begin{figure*}
\begin{center}
\psfig{width=\textwidth,file=rhoxyg53f1.eps.a,angle=0,clip=}
\caption{Density contours in the orbital plane during the dynamical
simulation of the black hole--neutron star binary with initial mass
ratio $q=0.31$ and $\Gamma=5/3$ (run D31). The orbital rotation is
counterclockwise. All contours are logarithmic and equally spaced
every 0.25 dex. Bold contours are plotted at $\log
\rho=-5,-4,-3,-2,-1$ (if present) in the units defined in
eq.~\ref{eq:defrhounit}. The thick black arcs bound the matter that
forms the core (see \S~\ref{method}). The time for each frame is given
in the units defined in eq.~\ref{eq:deftunit}.}
\label{rhog53}
\end{center}
\end{figure*}

\setcounter{figure}{2}
\begin{figure*}
\begin{center}
\psfig{width=\textwidth,file=rhoxyg53f2.eps.a,angle=0,clip=}
\caption{continued}
\end{center}
\end{figure*}

For every dynamical run, the decrease in binary separation leads to
Roche lobe overflow on an orbital time--scale. A stream of gas forms
at the inner Lagrange point, transferring matter from the neutron star
to the black hole. At the same time, the star is tidally stretched and
extends away from the black hole through the external Lagrange
point. We show in Figures~\ref{rhog2} and \ref{rhog53} density
contours in the orbital plane at various times during the simulation
for runs C31 and D31. For all other runs (C51, C31S, C20, D51, D31S
and D20) the plots are qualitatively similar. As the accretion stream
winds around the black hole, it collides with itself and forms a
torus, while the gas thrown out through the outer Lagrange point forms
a long tidal tail. For a the less compressible case ($\Gamma=2$, run
C31), the torus, as well as the tidal tail, are thinner, as one should
expect. The accretion torus that is formed around the black hole is
not initially azimuthally symmetric, but shows a double ring
structure, particularly for $\Gamma=2$ (see panels (d)--(h) in
Figure~\ref{rhog2}). This appears as the gas that passes through
periastron near the black hole overshoots the circular orbit that
would correspond to the specific angular momentum it contains, forming
an outer ring (see panels (c)--(d) in Figure~\ref{rhog2}). It then
falls back towards the black hole and encounters the rear of the
accretion stream. The subsequent collision tends to circularize the
orbit of the fluid, and also pushes it to the inner ring, closer to
the black hole (panels (d)--(e)). The structure of the outer ring
rotates slowly counterclockwise (with the initial orbital motion) as
the bulk of the tidally disrupted star (which produces the accretion
stream) continues orbital motion in the same direction, on the
opposite side of the black hole. At late times, the density contrast
between the rings drops (see below, Figure~\ref{rho70g2g53}(a) and
(c), and Figure~\ref{tailg2}), but nevertheless a hump remains in the
accretion disc, as there is still a visible stream feeding it from the
oposite side. This structure was clearly seen for $\Gamma=2.5$ in the
results presented in paper~III, and for the same reasons. It is
present as well for $\Gamma=5/3$ (see Figure~\ref{rhog53}), although
the distinction between having two rings and a hump is not as marked,
even as the disc is forming (panels (c)--(e)). This is due to the
higher compressibility of the material, and hence its tendency to
expand at low densities. At late times the disc is much more
azimuthally symmetric than for $\Gamma=2$ (see below
Figure~\ref{rho70g2g53}(b) and (d)).

Our implementation of gravitational radiation reaction is valid only
for circular orbits. Thus we monitor the eccentricity $e$ of the
orbital motion during the coalescence, to ensure that it remains small
before gravitatational radiation reaction is switched off. We compute
an estimate for $e$ by assuming that it is that of a binary system
with masses $M_{\rm BH}$ and $M_{\rm core}$, given by $e=\sqrt{1+2E
J^{2}/G^{2}\mu M_{\rm core}^{2}M_{\rm BH}^{2}}$, where $\mu=M_{\rm
core} M_{\rm BH}/(M_{\rm core}+M_{\rm BH})$, and $E$ and $J$ are the
mechanical energy and angular momentum of the orbital motion. During
the initial phase, $e\approx 0.05$, and close to the instant of
minimum binary separation (at $t\simeq 20$ for most runs) $e<0.1$. By
the time radiation reaction is switched off (at $t\simeq 30$, see
Table~\ref{parameters}), the eccentricity has increased somewhat, to
$e\simeq 0.2$. At this stage the mass ratio (between the core and the
black hole) has dropped enough so that the effects of including
radiation reaction are very small.

The separation between the centre of mass of the core and the black
hole is shown in Figure~\ref{rt} for all runs.
\begin{figure*}
\psfig{width=\textwidth,file=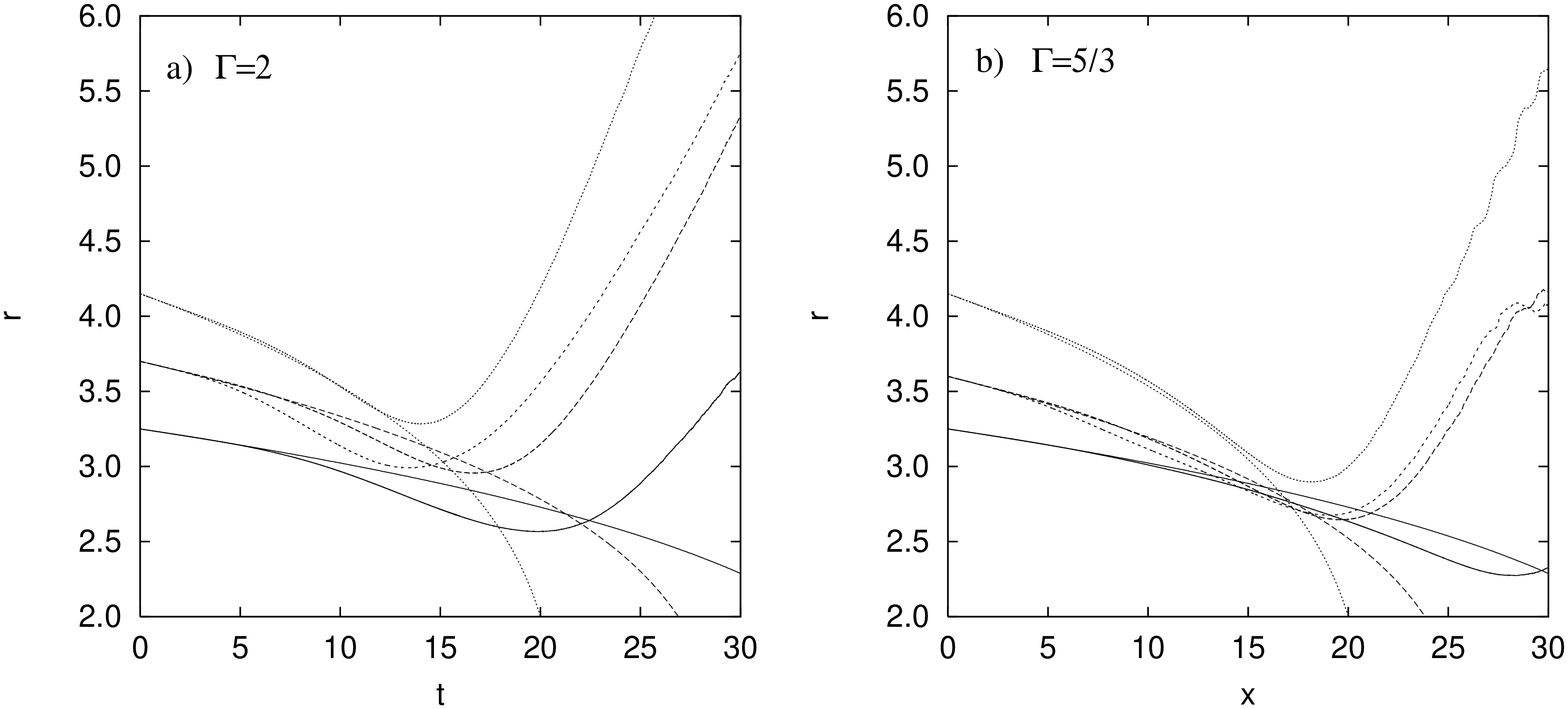,angle=0,clip=}
\caption{Separation between the black hole and the centre of mass of
the core (see \S~\ref{method}) for (a)~$\Gamma=2$ and (b)~$\Gamma=5/3$
(runs C50 and D50---solid lines; runs C31 and D31---long--dashed
lines; runs C31S and D31S---short--dashed lines; runs C20 and
D20--dotted lines). For $q=0.31$ there are two curves in each frame,
corresponding to runs initiated with a spherical polytrope and an
irrotational Roche--Riemann ellipsoid. In both cases, the one that
decays faster corresponds to the former condition. The monotonically
decaying curves correspond to point--mass binaries with the same
initial separation and mass ratio, evolving through gravitational wave
emission, computed in the quadrupole approximation.}
\label{rt}
\end{figure*}
Initially, it decreases at a rate consistent with that of a
point--mass binary, and subsequently does so at an even faster rate,
due to hydrodynamical effects. This is particularly important for high
mass ratios ($q=0.5$ and $q=0.31$). For $q=0.2$, the deviation is
smallest and almost negligible, until $t\approx 13-15$, depending on
the value of $\Gamma$. Once the initial mass transfer episode is under
way, the separation reaches a minimum and then increases, as the core
of the star is stretched and moves to a greater
separation. Qualitatively, the evolution resembles that of a stiff
equation of state (paper~III) except that now the point--mass
approximation for the orbital decay is valid for a longer time (at
smaller separations for a given value of $q$, compare for example the
case with $q=0.2$ in the two panels in Figure~\ref{rt}). This is
simply due to the fact that the stars are better approximated by point
masses as the adiabatic index is decreased. The point at which the
separation is at a minimum coincides with the maximum accretion rate
(see below). After this initial periastron passage, the star is
completely disrupted, and in every case the final configuration
consists of a massive accretion disc around the black
hole. Gravitational radiation reaction is switched off at $t\approx
30$ for all runs (see Table~\ref{parameters}) when the core mass drops
below 0.14~$M_{\odot}$.

\begin{figure*}
\psfig{width=\textwidth,file=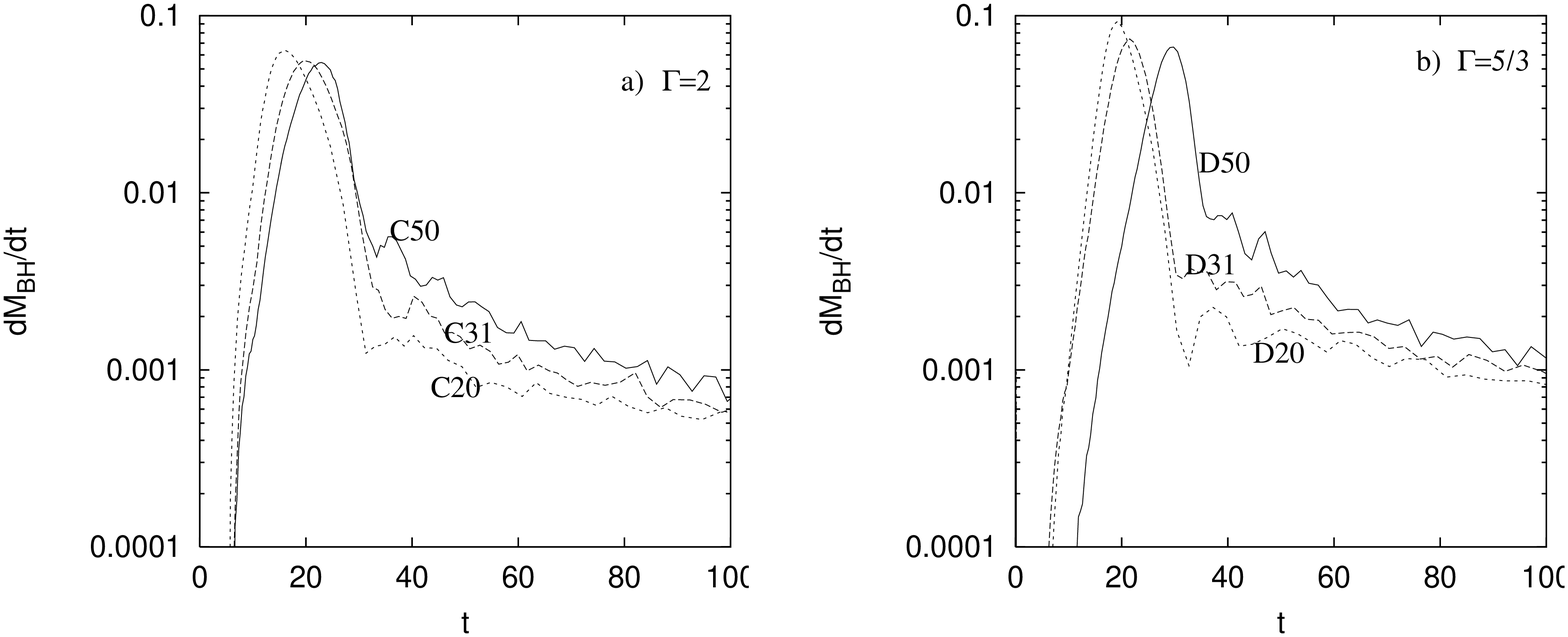,angle=0,clip=}
\caption{Mass accretion rate onto the black hole for (a) $\Gamma=2$
and (b) $\Gamma=5/3$. The curves are plotted only for $t<100$, at
later times there is little further evolution as $\dot{M}$ decreases
monotonically.}
\label{mdot}
\end{figure*}

As mentioned above, the material which moves away from the black hole
through the outer Lagrange point forms a long one--armed spiral in the
system. This structure is usually formed during a dynamical
coalescence (it is a two--armed spiral in the case of neutron star
mergers, with each star producing one arm, see e.g. RS94). The main
difference between the runs presented here and the case of low
compressibility studied in paper~III is that no clumps are formed, and
the distribution of matter remains smooth along the length of the tail
(as observed also in paper~II, RS92, RS95 for soft equations of
state).

We show in Figure~\ref{mdot} the accretion rates onto the black hole
for runs C50, C31, C20, D50, D31 and D20.  The maximum rates are
reached during the initial episode of Roche lobe overflow
($\dot{M}_{max}\approx 0.06-0.09$, equivalent to
$0.7-1.1$~$M_{\odot}$~ms$^{-1}$). They correspond mainly to matter
that is directly accreted by the hole from the mass transfer stream
coming from the neutron star.  As the accretion disc is formed around
the black hole, $\dot{M}$ gradually decreases, although there are
small oscillations. One can see in the curves that there are secondary
maxima in $\dot{M}$ at $t<60$ for all runs. This is due to the
circularization process of the orbits in the disc. The streams of
matter coming from the neutron star collide with themselves, and some
matter (along the inner edge of the stream) falls onto the black hole
with greater ease, giving rise to the quasi--periodic oscillations in
the accretion rate. This only occurs two or three times at most, and
at late times ($t\geq 100$) the accretion rate is decreasing
monotonically, showing the circularization of the orbits.  These peaks
are present in the runs shown in paper~III, but the lower resolution
used there makes it harder to appreciate them.

The peak accretion rates shown in Table~\ref{disks} are substantially
higher when $\Gamma=5/3$, by up to a factor of $\simeq 1.4$ for
$q=0.2$. This is one of many effects of the mass--radius relationship
that are observed during dynamical coalescences. For polytropes, $R
\propto M^{\Gamma-2/(3\Gamma-4)}$, so for $\Gamma=2$ the stellar
radius is unaffected by mass loss (or accretion), while for
$\Gamma=5/3$, $R \propto M^{-1/3}$ and thus the star will expand upon
overflowing its Roche lobe and losing mass. The decrease in separation
due to energy losses to gravitational waves is what initially brings
the star to the point of Roche lobe overflow. The star's reaction when
this happens then depends on the compressibility. When $\Gamma=2$, the
effect of tidal forces and gravitational wave back reaction are enough
to completely disrupt the star. For $\Gamma=5/3$, there is an added,
runaway effect, because since the star {\em expands} upon losing mass,
it further overflows its Roche lobe (and thus produces higher
accretion rates). This alone can de--stabilize the orbit and induce
coalescence, as observed for the case of a tidally locked system with
$\Gamma=5/3$, {\em without} gravitational wave back reaction (this was
done initially as a test and reported in paper~II). This also explains
why the total disc mass is lower for a given initial mass ratio for
$\Gamma=5/3$ (also given in Table~\ref{disks}, column 2).

The total angular momentum in the system decreases for two
reasons. First, there is a decay due to the emission of gravitational
waves (seen at early times before substantial mass transfer has taken
place), and second, much of the angular momentum of the accreted
matter is lost to the spin of the black hole. As stated above, when
accretion occurs we update the mass and momentum of the black hole so
as to ensure conservation of total mass and linear
momentum. Conservation of angular momentum then allows us to estimate
the degree to which the black hole is spun up as a result of
accretion, and we calculate its Kerr parameter $a=J_{\rm BH}c/G M_{\rm
BH}^{2}$ at the end of the simulation (we take $a=0$ at $t=0$). This
is shown in column 8 of Table~\ref{disks}. We note that our previously
published results for $a$ in papers~II and III contained an error
which we have now corrected. An explanation and the correct values are
given in the appendix. For a fixed adiabatic index, the black hole is
spun up to a greater degree (up to almost one half the maximum
rotation rate) at higher mass ratios, simply because it is less
massive. The higher rotation rates seen at lower $\Gamma$ reflect the
corresponding higher accretion rates (see above) and the fact that the
total accreted mass is greater.

The various energies in the system are shown in Figure~\ref{energies}
for runs C31 and D31.  The changes in mechanical energy seen at early
times are due to the back reaction of gravitational waves, while the
initial episode of mass transfer in the initial stages of the
coalescence is evident in the large changes that occur at $t\approx
15$. At later times, the variation is small and the curves show a
monotonical decay as the accretion discs become more azimuthally
symmetric.

The core of the neutron star moves away from the black hole for the
same reasons as outlined in paper~III for the case of a stiff equation
of state. In the case of conservative mass transfer (where the total
mass and orbital angular momentum $J$ are conserved), if the donor is
the less massive component, the binary separation will increase. The
system is not strictly conservative in this case, but the global
response is the same. The specific angular momentum in the core
increases as mass transfer proceeds, and this makes the separation
increase. The mass--radius relationship outlined above makes it
impossible for the system to survive as a stable binary, as was the
case for $\Gamma=3$ (paper~III). As soon as the star overflows its
Roche lobe, catastrophic mass transfer ensues, completely disrupting
the star. The two dominant effects as far as the orbital evolution is
concerned are the gravitational wave emission (and the accompanying
loss of angular momentum) and mass transfer. In the cases shown here,
(where the neutron star expands or maintains a constant radius upon
losing mass), both effects lead to complete tidal disruption of the
star on an orbital time--scale. For a stiff equation of state they
tended to drive the system in opposite directions, with angular
momentum losses making the separation decrease while mass transfer
increased it. The outcome in that case was {\em episodic} mass
transfer, and the frequent formation of accretion discs when the star
was disrupted. Thus, in the present case as well, {\em stable} mass
transfer from the neutron star is impossible, and the final
configuration consists of a massive accretion disc around the black
hole.

\begin{figure}
\psfig{width=7.5cm,file=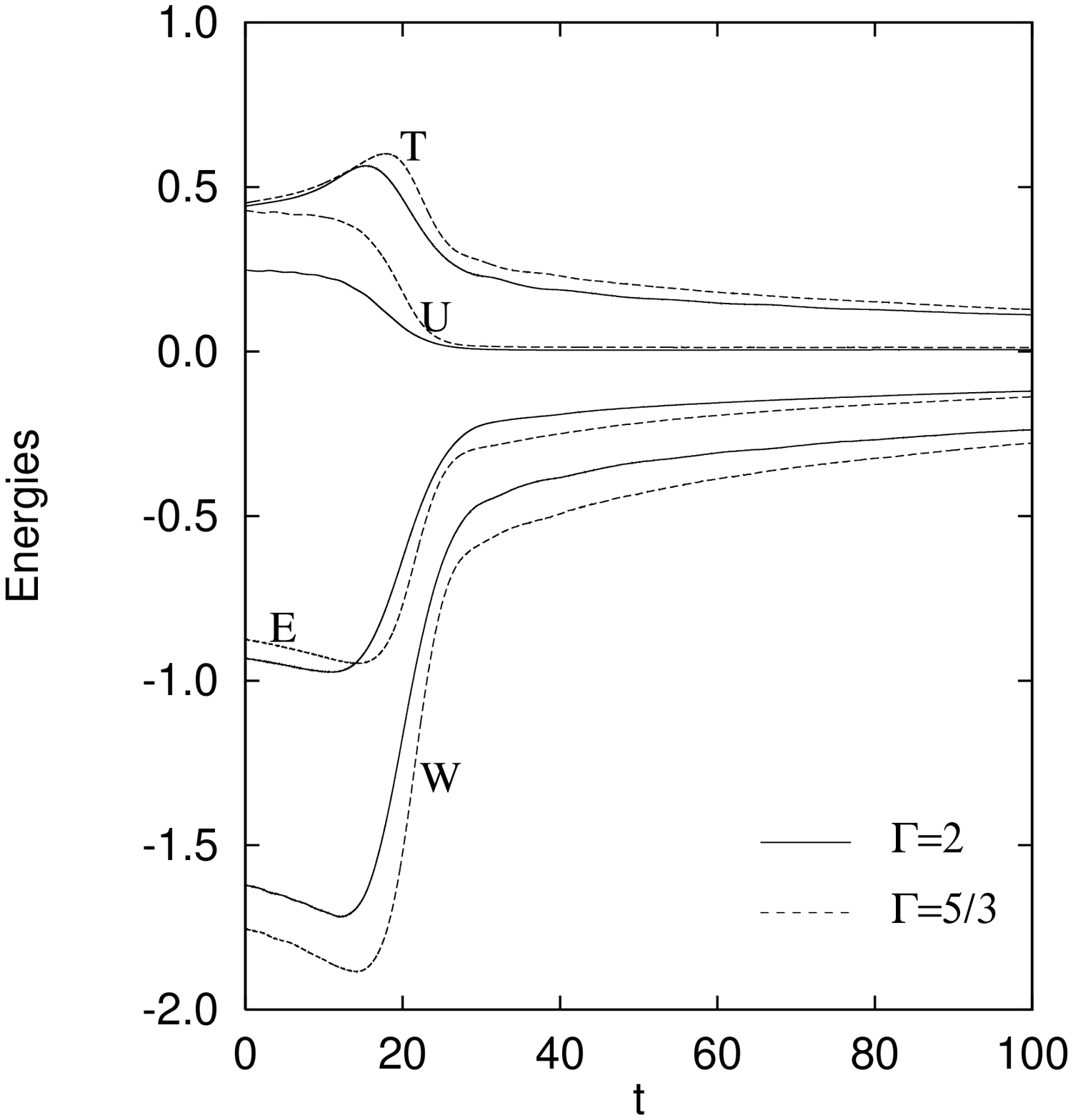,angle=0,clip=}
\caption{Energies in the system as a funcion of time for $q=0.31$
(runs C31 and D31). The kinetic (T), total internal (U), gravitational
potential (W) and total (E) energies are given in units of $3.8 \times
10^{53}$~erg. }
\label{energies}
\end{figure}

\subsection{Accretion disc structure}

In Table~\ref{disks} we show several parameters related to the
accretion structure around the black hole at the end of the
simulation. The disc masses are computed as before (papers~II and
III), by taking into account the mass which has specific angular
momentum $j>\sqrt{6}GM_{total}/c$, so that it will remain in orbit
around the black hole. This means that only a fraction $f$ of the gas
mass left in the system at the end of the simulation is considered for
the accretion tori (usually around 70\%, the specific value of $f$ for
each run is given in the third column of Table~\ref{disks}). The disc
masses are lower at lower $\Gamma$ (as mentioned above) and at a lower
initial mass ratio, but there is always at least 0.1~solar masses left
in a debris torus. From the final accretion rate (comparable in every
case) and the disc mass at $t=t_{f}$ we infer a rough estimate of the
lifetime of the disc, which is beteen 30 and 70~ms. Since it is
viscosity that drives the evolution of the disc at late times, and
hence in this case a purely numerical effect, these values should be
taken only as a guideline. The true lifetime of the disc depends on
the redistribution of angular momentum through viscous effects and
possibly dynamical instabilites. The former would probably make for a
longer--lived disc, while the latter would tend to act in the opposite
direction.

The double ring structure mentioned above gradually disappears as the
density contrast between the rings drops, and the disc becomes more
azimuthally symmetric as the simulation progresses. By $t=t_{f}$ it is
meaningful to take azimuthal averages of quantities such as the
density, internal energy and specific angular momentum in the
disc. These are shown in Figure~\ref{diskprof} for runs C31 and D31
(the corresponding plots for the remaining runs are quite similar).
The density has a maximum at a characteristic distance $r_{0}$, which
is between 50 and 70~km. (Another estimate of the lifetime of the disc
can be obtained by evaluating $|r_{0}/v_{r}|$, where $v_{r}$ is the
radial (inward) velocity of the fluid at $r_{0}$. The resulting
lifetimes $\tau_{infall}$ are of the same order as those shown in
Table~\ref{disks}.) The specific internal energies are maximum in the
inner regions of the discs, and the profiles flatten out at $r\approx
2r_{0}$, at about $u/1000=2\times10^{-5}$, equivalent to
$3\times10^{18}$~erg~g$^{-1}$, or $3$~MeV~nucleon$^{-1}$. The rotation
curves are subkeplerian, indicating that pressure support is
important. The slight increase in the curves of specific angular
momentum seen in Figure~\ref{diskprof}b at $r\approx 10$ (particularly
for $\Gamma=2$) are due to the persistent outer ring in the accretion
disc (see Figure~\ref{rho70g2g53}). The slow drop in specific angular
momentum for $r>10$ marks the edge of the accretion disc at
$t=t_{f}$. We note that in the inner regions, where the orbits have
been circularized through the dynamical process mentioned above, the
rotation curves are always close to, and below the Keplerian value.
The torus does {\em not} have a constant distribution of specific
angular momentum $j$, even immediately after being formed.

Since one of our main motivations for this line of work has been the
production of cosmological gamma ray bursts from theses systems, we
show as usual the distribution of matter along the rotation axis. This
is the region where the densities are lowest, and from which a
relativistic fireball could possibly emerge from the system and
produce a GRB. This could possibly be powered by neutrino emission
from the disc and subsequent pair production and expansion, or through
the Blandford \& Znajek \shortcite{blandford77} mechanism, by tapping
the energy stored in the spin of the black hole (see
Table~\ref{disks}). In order for this to occur, the baryon loading
must be small (on the order of $10^{-5}M_{\odot}$), so that the
expansion can occur at the required Lorentz factors $\Gamma \geq
10^{2}$ \cite{rees92,rees93}. In Figure~\ref{mtheta} we show the
baryon contamination along the rotation axis by plotting the amount of
mass enclosed in a cone with opening angle $\Delta \theta$ above and
below the black hole and along the rotation axis (see also the last
three columns in Table~\ref{disks}). Clearly, only modest collimation
is required (of about $10^{\circ}$) to stay below the
$10^{-5}M_{\odot}$ threshold mentioned above. This is a fact that has
become clearer in our simulations as we have increased the resolution,
from 8,000~SPH particles for the majority of the runs shown in paper~I
to the 80,000 particles used for the simulations shown here (in
increasing the number of particles by an order of magnitude, the
spatial resolution increases by a factor of $\simeq 2.15$).

\begin{figure*}
\psfig{width=\textwidth,file=rho70g2g53.eps.a,angle=0,clip=}
\caption{Density contour plots at $t=t_{f}$ for runs C31~(a,b) and
D31~(c,d) in: (a,c) the orbital plane; (b,d) the meridional plane
shown by the black line in panels (a,c). All contours are logarithmic
and equally spaced every 0.25 dex. Bold contours are plotted at $\log
\rho=-6,-5,-4$ (if present) in the units defined in
eq.~\ref{eq:defrhounit}.}
\label{rho70g2g53}
\end{figure*}

\begin{table*}
 \caption{Accretion disc structure. In the last three columns,
$\theta_{-n}$ is the half--angle of a cone above the black hole and
along the rotation axis of the binary that contains a mass
$M=10^{-n}$. The mass is given in units of 1.4~$M_{\odot}$, and time
is measured in the units defined in equation~\ref{eq:deftunit}.}
 \label{disks}
 \begin{tabular}{@{}lcccccccccc}
  Run & $M_{disc}$ & $f$ & $\dot{M}_{max}$    
	& $\dot{M}_{final}$
        & $M_{ejected}$
        & $\tau_{disc}$ & $J_{\rm BH}c/G M_{\rm BH}^{2}$ &
        $\theta_{-3}$
        & $\theta_{-4}$ & $\theta_{-5}$ \\
  C50   & 0.181 & 0.71 & 0.054 & 3$\cdot 10^{-4}$
        & 0.48$\cdot 10^{-3}$ & 603 
        & 0.334 & 41 & 25 & 15\\
  C31   & 0.172 & 0.71 & 0.057 & 3$\cdot 10^{-4}$
	 & 10.20$\cdot 10^{-3}$ & 573
        & 0.234 & 46 & 30 & 18\\
  C31S   & 0.179 & 0.73 & 0.053 & 3$\cdot 10^{-4}$
	 & 11.51$\cdot 10^{-3}$ & 596
        & 0.232 & 48 & 30 & 18\\
  C20   & 0.138 & 0.59 & 0.065 & 3$\cdot 10^{-4}$
	 & 6.97$\cdot 10^{-3}$ & 460
        & 0.162 & 52 & 38 & 25\\
  D50   & 0.159 & 0.19 & 0.067 & 3$\cdot 10^{-4}$
	 & 0.21$\cdot 10^{-4}$ & 530 
        & 0.343 & 35 & 19 & 12\\
  D31   & 0.141 & 0.63 & 0.074 & 3$\cdot 10^{-4}$
	 & 0.80$\cdot 10^{-4}$ & 470
        & 0.242 & 43 & 26 & 12\\
  D31S   & 0.144 & 0.64 & 0.072 & 3$\cdot 10^{-4}$
	 & 2.14$\cdot 10^{-4}$ & 480
        & 0.241 & 46 & 26 & 13\\
  D20   & 0.086 & 0.50 & 0.094 & 3$\cdot 10^{-4}$
	 & 0.05$\cdot 10^{-4}$ & 286
        & 0.172 & 51 & 35 & 22\\

 \end{tabular}

\end{table*}

\subsection{Ejected mass} \label{ejected}

During the initial encounter, a tidal tail of material stripped from
the neutron star is formed in every dynamical run. This tail has been
observed before for a stiff equation of state (paper~III), where it
persists as a well--defined structure throughout the simulations. For
the cases studied here, it is present for $\Gamma=2$ (see
Figure~\ref{tailg2}), but essentially disappears as a coherent
structure at late times for $\Gamma=5/3$, as the density drops and the
tail expands. We have calculated the amount of dynamically ejected
mass for every run as before, computing the total mechanical energy
(kinetic+gravitational potential) of the fluid, and counting as
ejected that mass for which it has a positive value at
$t=t_{f}$. There are two distinct categories of ejected mass during
the simulation. The first (type~I) corresponds to matter {\em
dynamically} ejected from the system, and can be found in the orbital
plane, at the tips of the tidal tails formed during the disruption of
the star at early times ($t<30-40$, see panels (b)--(d) in
Figures~\ref{rhog2} and~\ref{rhog53}, and Figure~\ref{tailg2}). The
second (type~II) comes from the surface of the accretion disc, and is
ejected from the system at later times ($t>70$). Ejected matter of
type~II only appears in a significant amount for the runs with
$\Gamma=5/3$, and is mainly of numerical origin, as it is the equation
of state that models more compressible gas. This means that the fluid
expands to occupy a larger volume than for less compressible equations
of state. Thus for a given number of SPH particles, the spatial
resolution is lower (i.e. the smoothing length $h$ is larger),
particularly at the edge of the matter distribution, and the effects
of heating due to the artificial viscosity can be more pronounced.  It
was not mentioned in paper~III simply because no resolvable amount of
mass was ejected in this fashion. For $\Gamma=2$ it amounts to only a
tiny fraction of the total ejected mass (and a handful of
particles). For $\Gamma=5/3$ however, this is no longer true. In fact,
most of the ejected matter is type~II in this case. We have not
counted it in the values tabulated in the sixth column of
Table~\ref{disks}, keeping only type~I ejected matter. Including the
internal energy $u$ of the fluid does not alter the results given,
since the gas coming from the tips of the tidal tails has not been
subjected to strong compression and heating, as it was never part of
the accretion torus around the black hole.

\begin{figure*}
\psfig{width=\textwidth,file=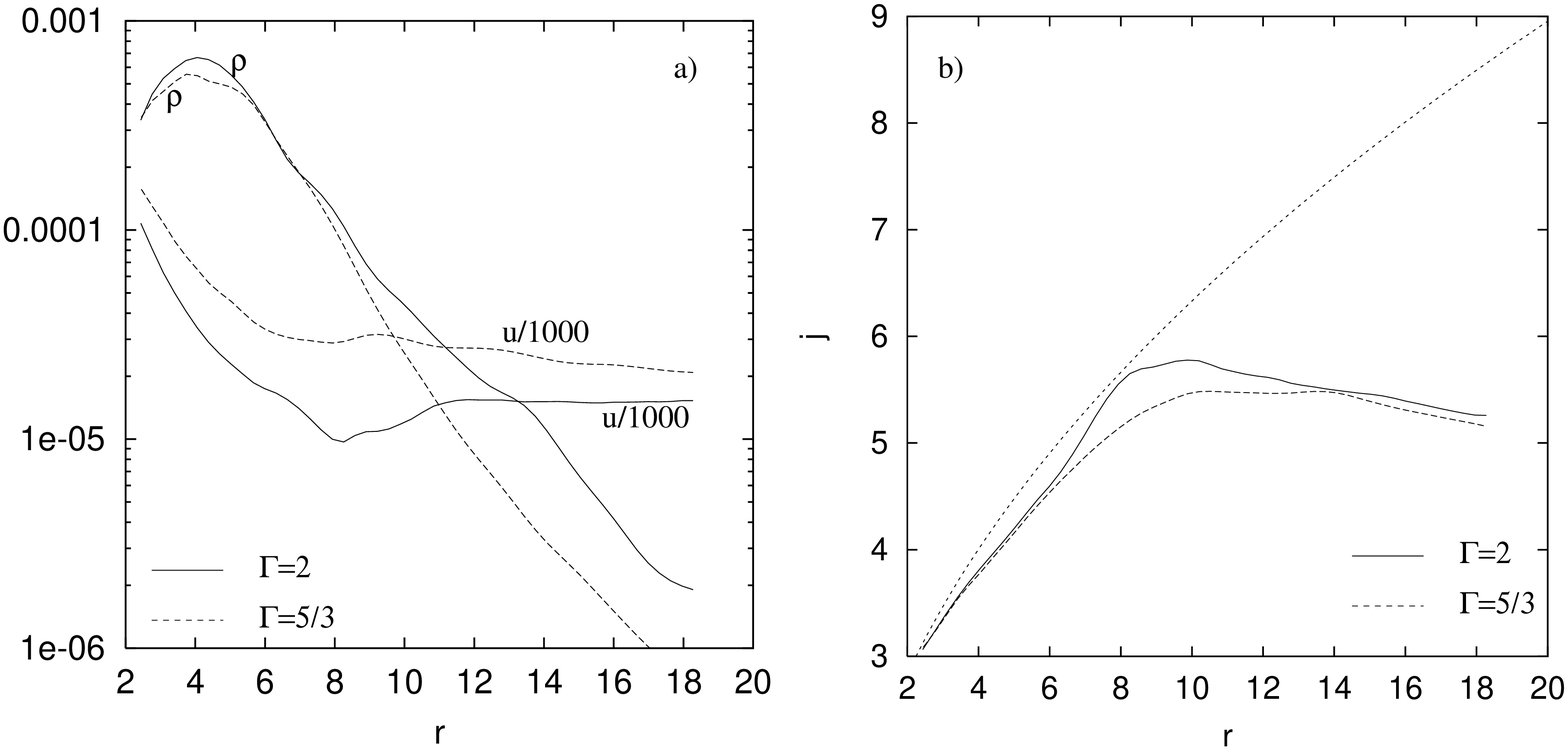,angle=0,clip=}
\caption{(a) Azimuthally averaged profiles for the density $\rho$ and
internal energy $u$ ($u/1000$ is plotted) for runs C31 ($\Gamma=2$)
and D31 ($\Gamma=5/3$) in the equatorial plane at $t=t_{f}$. (b)
Specific angular momentum $j$ in the equatorial plane for the same
runs as in (a). The monotonically increasing curve corresponds to that
of a Keplerian accretion disc around a black hole of the same mass
(the mass of the black hole at $t=t_{f}$ for runs C31 and D31 differs
by less than 1 per cent).}
\label{diskprof}
\end{figure*}

\begin{figure*}
\psfig{width=\textwidth,file=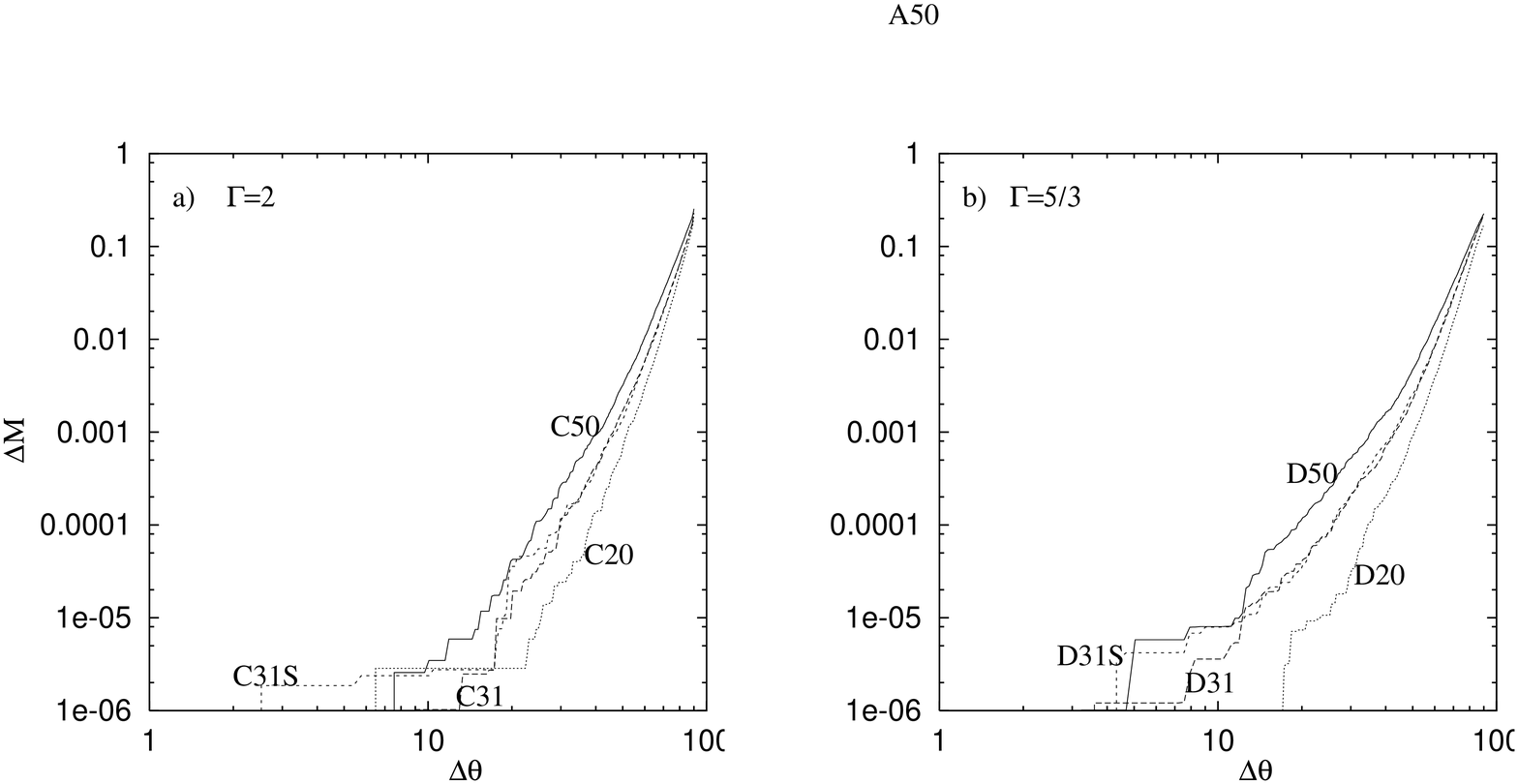,angle=0,clip=}
\caption{Enclosed mass for all runs as a function of half--angle
$\Delta \theta$ (measured from the rotation axis in degrees) for (a)
$\Gamma=2$ and (b) $\Gamma=5/3$ at $t=t_{f}$.}
\label{mtheta}
\end{figure*}

For the softer equation of state, mass ejection is strongly
suppressed, by approximately two orders of magnitude. This effect is
so strong that for run D20 only about 20 SPH particles leave the
system. This is similar to what was observed by Rosswog et
al. \shortcite{rosswog99} in the case of double neutron star
mergers. The underlying reason is that as the adiabatic index is
lowered, the star becomes more centrally condensed, and thus the
gravitational potential well becomes progressively deeper. For
polytropes, the gravitational potential and the density are related by
$\Phi=-K_{\Gamma}\Gamma/(\Gamma-1)\rho^{\Gamma-1}$ where $K_{\Gamma}$
depends on the value of $\Gamma$. This gives
$\Phi_{c}(\Gamma=2)=0.74\Phi_{c}(\Gamma=5/3)$, at a constant stellar
mass and radius. This alone makes it more difficult to extract gas
from the stellar potential well, through the gravitational interaction
with the black hole during coalescence and eject it from the
system. In all simulations, we see that the gas that is dynamically
ejected (type~I) comes from the surface layers of the star. So if the
potential well is deeper, less matter is available for this sort of
ejection, all other things being equal. There are at least three more
effects that enhance this result and tend to decrease the amount of
ejected mass at higher compressibilities. The first is that more
violent events (as measured for example by the departure from
point--mass behavior at small separations, see Figure~\ref{rt}) tend
to eject more matter. Since these deviations are driven precisely by
hydrodynamical effects, their influence is reduced at low
$\Gamma$. Second, as pointed out above, the ejected matter comes from
the surface layers of the star, and thus from regions that are at
lower density at low values of $\Gamma$, making for less total matter
available for ejection. Third, as can be seen in Figure~\ref{jr},
there is less total angular momentum in the system as $\Gamma$ is
decreased (also due to a greater degree of central condensation in the
star), and so it will be more difficult for matter to escape the
system in that case.

The combination of the effects mentioned above makes for a dramatic
drop in the value of $M_{ejected}$ given in Table~\ref{disks} as a
function of $\Gamma$. The transition is sharp, due to the gradual
increase in central condensation of the star, and in particular to the
qualitative change in the mass--radius relationship that occurs at
$\Gamma=2$.

\subsection{Emission of gravitational waves}

The waveforms and luminosities are calculated in the quadrupole
approximation from the values of the reduced moment of inertia tensor,
and its time derivatives, see e.g. Finn \shortcite{finn} and RS92. One
polarization of the waveforms is shown in Figure~\ref{gwpol} for runs
C31 and D31, compared with the result for point masses decaying in the
same approximation, and the corresponding luminosities are plotted in
Figure~\ref{gwlum}. The results are very similar for the dynamical
runs with a different initial mass ratio (runs C50, D50, C20 and D20).

\begin{figure}
\psfig{width=7.5cm,file=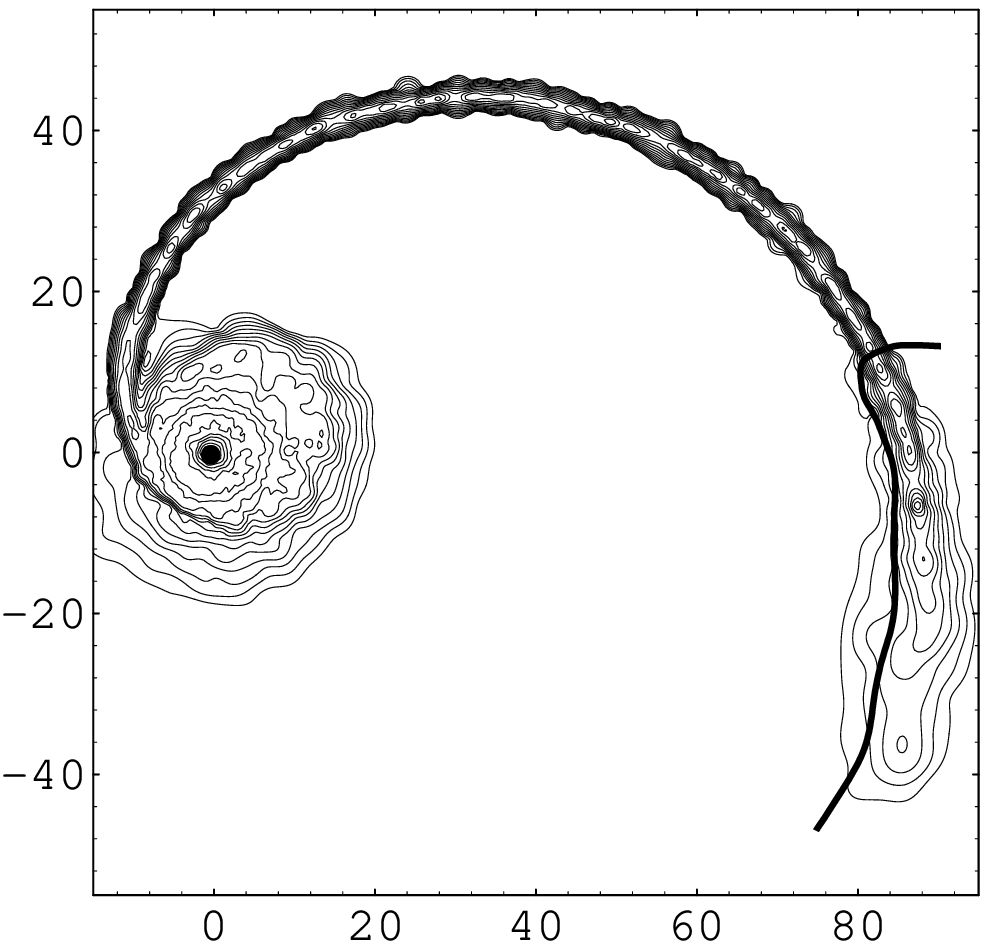,angle=0,clip=}
\caption{Density contours in the orbital plane at $t=t_{f}$ for run
D31. All contours are logarithmic and equally spaced every 0.25
dex. Bold contours are plotted at $\log \rho=-6,-5,-4$ in the
units defined in eq.~\ref{eq:defrhounit}. The thick black line across
the tidal tail divides the matter that is bound to the black hole from
that which is on outbound trajectories.}
\label{tailg2}
\end{figure}

The tidal disruption of the neutron star in every run (irrespective of
the mass ratio) after the first episode of mass transfer following
periastron passage, makes the amplitudes of the waveforms drop
abruptly, and practically to zero, as the accretion torus is formed
and becomes ever more azimuthally symmetric. Upper bounds for the
final amplitude (at $t=t_{f}$) are shown in Table~\ref{waves}, where
we show the maximum and final amplitudes for the waveforms, the peak
luminosity and the total energy radiated away by the system, and the
efficiency of gravitational wave emission $\epsilon=\Delta E/M_{\rm
total}c^{2}$. For reference, $L_{max}=1$ (in the units given in the
table) corresponds to $3.036\times 10^{55}$~erg~s$^{-1}$and $\Delta
E=10$ is equivalent to $3.48\times 10^{52}$~erg. The one--armed spiral
arms formed during the coalescence (see Figure~\ref{tailg2}) do not
contain enough mass to alter the waveforms significantly
($M_{tail}\approx0.05$). Since the total mass of the system is not the
same for each run, but increases as the mass ratio is decreased, the
peak amplitudes in the waveforms (as well as the peak luminositites)
are higher as well for lower $q$ (at a fixed value of $\Gamma$). At a
fixed value of the initial mass ratio, however, one can observe the
effect of using a different adiabatic index clearly. At higher
compressibility (i.e. lower $\Gamma$), the maximum amplitudes, peak
luminosities, the total energy release in gravitational waves and the
efficiency of this emission are all higher (see
Table~\ref{waves}). The reason for all these trends is the same: the
higher the compressibility, the more centrally condensed the star
is. For $\Gamma=5/3$, $\rho_{c}/\overline{\rho}=5.99$, while for
$\Gamma=2$, $\rho_{c}/\overline{\rho}=3.29$ ($\rho_{c}$ is the central
density of the star and $\overline{\rho}$ is its average
density). Thus, it resembles a point mass to a greater degree in the
case with $\Gamma=5/3$ than if $\Gamma=2$. It is precisely the
hydrodynamical effects associated with the star {\em not} being a
point mass that are driving the waveforms and luminosities away from
the point--mass result and making them decay. One can also see in
Figure~\ref{gwpol} that for $\Gamma=5/3$, the waveform takes longer to
begin the decay, and stays close to the point--mass result for a
longer time.

\begin{figure*}
\psfig{width=\textwidth,file=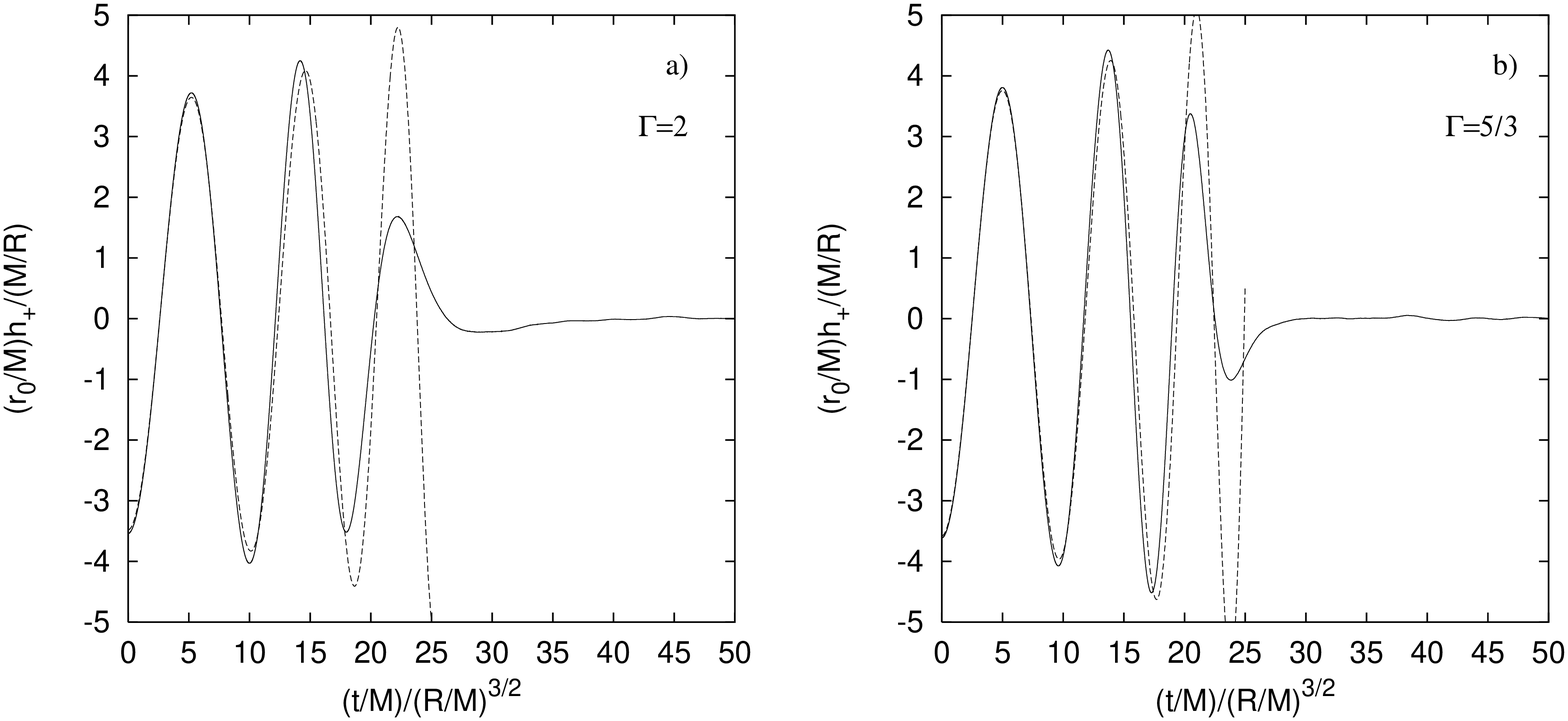,angle=0,clip=}
\caption{Gravitational radiation waveforms (one polarization is shown)
seen at a distance $r_{0}$ away from the system along the rotation
axis for runs C31 (a), and D31 (b). The dashed lines show the
corresponding curves for a point--mass binary with the same initial
mass ratio and separation, decaying in the quadrupole
approximation. All quantities are given in geometrized units such that
$G=c=1$.}
\label{gwpol}
\end{figure*}

\begin{figure}
\psfig{width=7.5cm,file=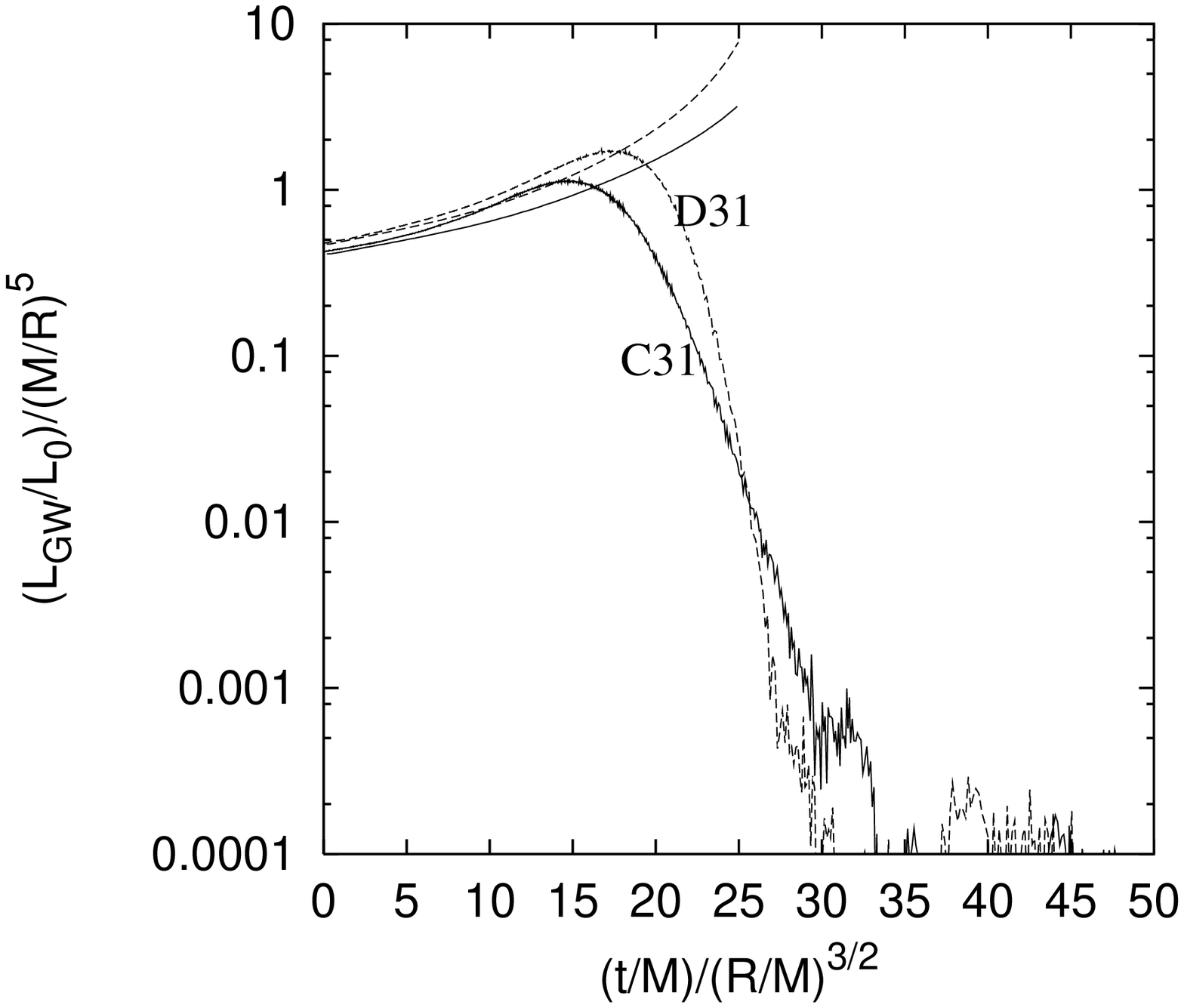,angle=0,clip=}
\caption{Gravitational radiation luminosity for the same runs as shown
in Figure~\ref{gwpol} (solid lines, run C31; dashed lines, run
D31). The monotonically increasing curves show the corresponding
result for a point--mass binary with the same initial mass ratio and
separation, decaying in the quadrupole approximation. All quantities
are given in geometrized units such that $G=c=1$
($L_{0}=c^{5}/G=3.64\times 10^{59}$~erg~s$^{-1}$).}
\label{gwlum}
\end{figure}

\begin{table*}
 \caption{Gravitational radiation. All quantities are given
in geometrized units such that $G=c=1$, and $L_{0}=c^{5}/G=3.64\times
10^{59}$~erg~s$^{-1}$.}
 \label{waves}
 \begin{tabular}{@{}lcccccr}
  Run & $(r_{0}R/M_{\rm NS}^{2})h_{max}$
   	& $(r_{0}R/M_{\rm NS}^{2})h_{final}$
        & $(R/M_{\rm NS})^{5}(L_{max}/L_{0})$
        & $(R^{7/2}/M_{\rm NS}^{9/2})\Delta E_{GW}$ & $\epsilon$ \\
      &      &      &      &      \\
  C50   & 3.00 & $\leq$ 0.01 & 0.58 & 8.95 & 4.15$\cdot10^{-3}$ \\ 
  C31   & 4.25 & $\leq$ 0.01 & 1.13 & 15.38 & 5.07$\cdot10^{-3}$ \\
  C31S  & 4.27 & $\leq$ 0.01 & 1.15 & 13.79 & 4.55$\cdot10^{-3}$ \\ 
  C20   & 5.80 & $\leq$ 0.01 & 2.10 & 24.95 & 5.79$\cdot10^{-3}$ \\ 
  D50   & 3.19 & $\leq$ 0.01 & 0.87 & 14.14 & 6.57$\cdot10^{-3}$ \\ 
  D31   & 4.55 & $\leq$ 0.01 & 1.71 & 22.29 & 7.35$\cdot10^{-3}$ \\ 
  D31S  & 4.49 & $\leq$ 0.01 & 1.58 & 21.86 & 7.21$\cdot10^{-3}$ \\ 
  D20   & 6.32 & $\leq$ 0.01 & 3.43 & 38.00 & 8.82$\cdot10^{-3}$ \\ 

 \end{tabular}

\end{table*}

\section{Influence of initial conditions on the dynamical evolution 
of the system} \label{sphere}

\begin{figure*}
\psfig{width=\textwidth,file=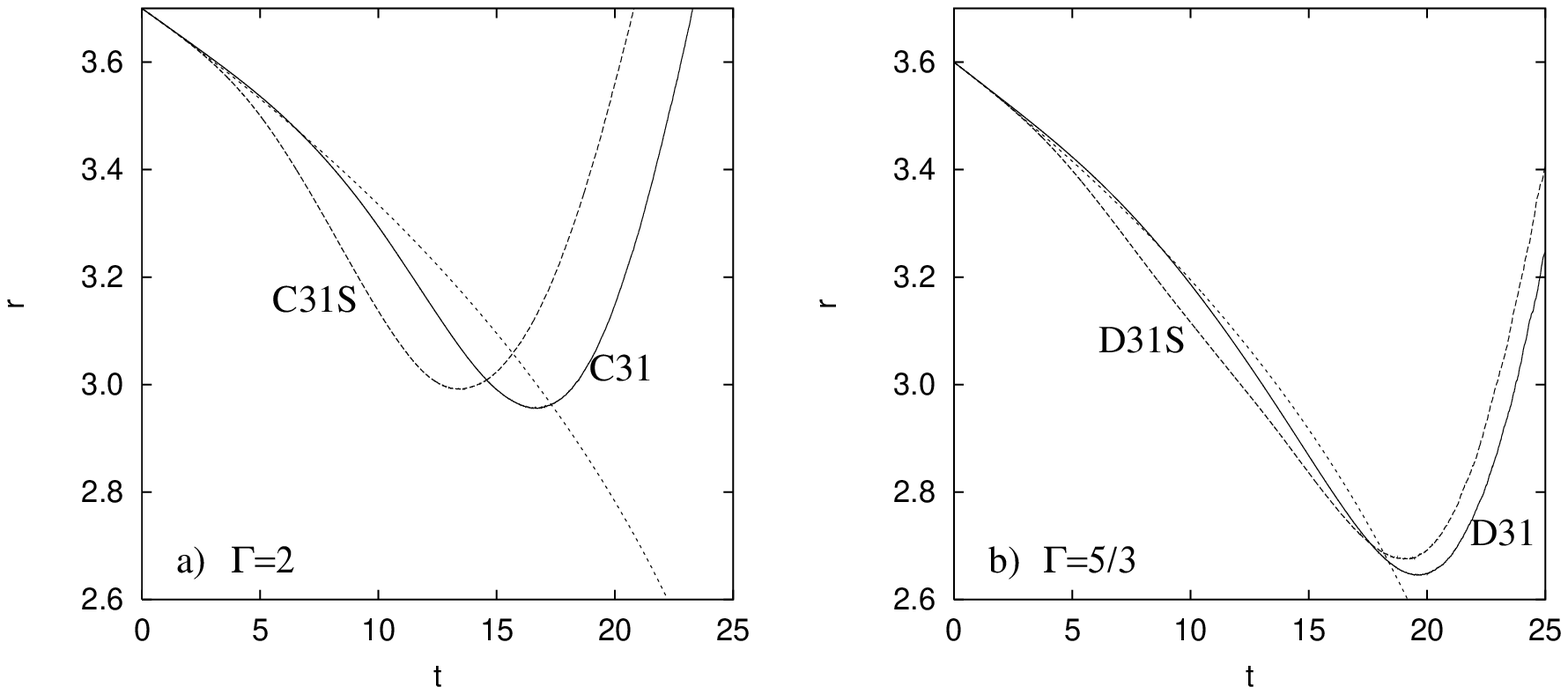,angle=0,clip=}
\caption{Separation between the black hole and the centre of mass of
the neutron star core as a function of time for (a) runs C31 and C31S
and (b) runs D31 and D31S. The monotonically decaying lines in each
frame are the result for a point--mass binary decaying through
gravitational wave emission, in the quadrupole approximation.}
\label{rtsph}
\end{figure*}

As for the results we presented in paper~III, there are two dynamical
runs that have used a spherical star as an initial condition, instead
of an irrotational Roche--Riemann ellipsoid. Both have an initial mass
ratio $q=0.31$, one for $\Gamma=2$ (run C31S) and one for $\Gamma=5/3$
(run D31S). The initial separation $r_{i}$ is the same as for runs C31
and D31. The initial orbital angular velocity $\Omega$ is that for
point--mass binaries, given that the axis ratios are
$a_{3}/a_{1}=a_{2}/a_{1}=1$. The purpose of these runs is to explore
the effect of using initial conditions that are far from equilibrium
for the calculations of dynamical coalescence. Since we have already
perfomed this type of run for a stiff equation of state in paper~III,
we can also gauge how strong the effects are as a function of the
compressibility. We remind the reader that, even if an irrotational
Roche--Riemann ellipsoid is a better approximation to the true
configuration of the system before coalescence than a spherical star,
it is not a {\em self--consistent} solution, since there are no true
equilibrium configurations for such a system. This is simply because
the emission of gravitational radiation is always present, and alters
the binary separation continuously. A tidal lag angle is always
present in the binary, because the bulge on the surface of the neutron
star cannot adjust to the changing gravitational potential
instantaneously. This angle remains small at large separations, but
can become quite large (on the order of $10^{\circ}$, see also Lai,
Rasio \& Shapiro~1994) just prior to coalescence, when the emission of
gravitational waves makes the potential change even faster. This
aspect of the coalescence process is greatly influenced by the
stiffness of the equation of state.

The strongest effect using a spherical star has on the dynamical
coalescence is due to the response of the star to the instantaneous
appearance of the gravitational field of the black hole at $t=0$. A
tidal bulge forms, along the line joining the two centres of mass. The
deformed star has a greater total energy $W_{self}+U$ (the internal
energy decreases, see Figure~\ref{utsph}, but the star is less bound
by gravity, and the net effect is to increase the energy), which is
taken in part from the orbital motion, and so the subsequent decay of
the orbit is faster than for runs C31 and D31 (see
Figure~\ref{rtsph}). The appearance of the tidal bulge also induces
radial oscillations in the star, which can be seen in the variations
of the total internal energy $U$, plotted in Figure~\ref{utsph} at
early times (compare also with the same curves for a stiff equation of
state, in Figure~14b of paper~III). The oscillations are always
present, but they are somewhat smaller for the runs initiated with
ellipsoids.

We shall focus on the results for $\Gamma=2$ for the following
discussion. There are slight variations if $\Gamma=5/3$ that we will
make clear below.  As mentioned above, the separation initially
decreases faster for run C31S than for run C31. However, the {\em
minimum} separation $r_{min}$ is slighty greater for run C31S (see
Figure~\ref{rtsph}), in contrast to the results shown in
paper~III. This is again because of the response of the neutron star
to mass loss. By expanding and overflowing its Roche lobe further
after the initial onset of mass loss, the encounter develops faster,
and the stellar core is pushed out to a larger binary separation
before approaching the black hole any further. This makes the peak
accretion rate lower, the final disc mass higher, and the Kerr
parameter of the black hole at $t=t_{f}$ marginally lower (see
Table~\ref{disks}). The gravitational radiation signal is also
affected by the initial condition, as can be seen in
Table~\ref{waves}. The faster orbital decay gives a higher peak
amplitude and luminosity (these quantities depend on the second and
third time derivatives of the moment of inertia respectively), but a
less energetic and efficient event, because it is more brief.

For $\Gamma=5/3$, the effect on the disc parameters given in
Table~\ref{disks} is the same as for $\Gamma=2$. However, there are
qualitative differences in the way the gravitational radiation signal
is affected.  The energetics and efficiency of the events vary in the
same way for runs D31 and D31S than for runs C31 and C31S, but the
trends are reversed as far as the peak amplitudes and luminosities are
concerned. The reason for this is that there are two important factors
determining the amplitude (and hence luminosities) of the
gravitational radiation waveform: the time derivatives of the moment
of inertia, and the mass ratio and separation. Inspection of
Figure~\ref{rtsph}b reveals that the differences between runs D31 and
D31S are small indeed. Essentially, the decay is accelerated by using
a spherical star, but not nearly fast enough to compensate for the
fact that the system attains a greater minimum separation. Thus the
peak amplitudes and luminosities are lower for run D31S than for run
D31.

\begin{figure}
\psfig{width=7.5cm,file=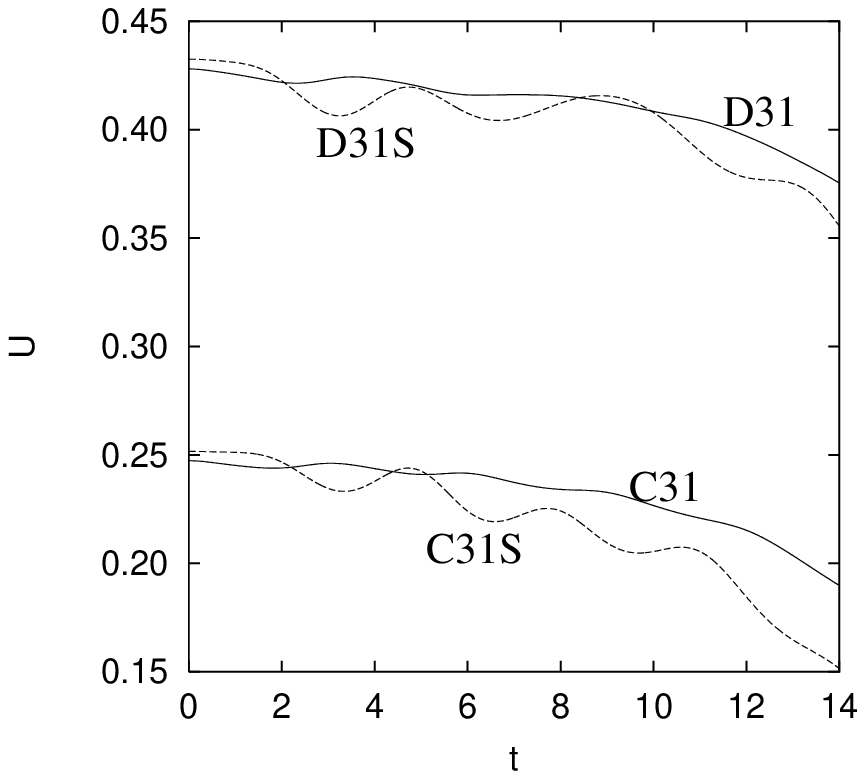,angle=0,clip=}
\caption{Total internal energy $U$ of the neutron star as a function
of time for runs C31, C31S, D31 and D31S.}
\label{utsph}
\end{figure}

\section{Summary, discussion and conclusions} \label{discussion}

We have presented the results of three--dimensional dynamical
simulations of the coalescence of a black hole with a neutron star,
using Smooth Particle Hydrodynamics. The black hole is modeled as a
Newtonian point mass with an absorbing boundary at the Schwarzschild
radius $r_{Sch}=2GM_{BH}/c^{2}$, and the neutron star is taken as a
cold polytrope with adiabatic index $\Gamma=2$ and $\Gamma=5/3$. The
spatial resolution of the results presented here is the highest we
have used to date, with $N\simeq 80,000$~SPH particles modeling the
initial neutron star. Dynamical runs with initial mass ratios ranging
from $q=M_{NS}/M_{BH}=0.5$ to $q=0.2$ were performed. Given that tidal
locking is not expected in these systems \cite{bildsten,kochanek}, we
have used initial conditions that correspond to irrotational binaries
in equilibrium, approximating the neutron star as a compressible
tri--axial ellipsoid, following the method of LRSb. The dynamical
simulations are begun when the system is on the verge of initiating
mass transfer, and followed for approximately 23~ms.

The binary separation decreases as a result of angular momentum losses
to gravitational radiation, and the neutron star overflows its Roche
lobe within one orbital period after the dynamical simulation is
started. Irrespective of the initial mass ratio and of the value of
the adiabatic index, this mass transfer episode leads to complete
tidal disruption of the star on an orbital time--scale. A massive
accretion disc forms around the black hole, containing a few tenths of
a solar mass (see Table~\ref{disks}). A single spiral arm appears,
from material moving through the outer Lagrange point, farthest from
the black hole. Initially, the accretion torus has a complicated
structure, with a double ring present (see Figures~\ref{rhog2} and
\ref{rhog53}), as the accretion stream collides with itself and
circularizes the orbits of the fluid in the disc. As the simulation
progresses, the disc becomes more and more azimuthally symmetric.  The
peak densities and specific internal energies in the discs at the end
of the simulations are on the order of $10^{11}$g~cm$^{-3}$ and
$10^{19}$~erg~g$^{-1}$ respectively (or about
10~MeV~nucleon$^{-1}$). All discs have a low degree of baryon
contamination along the rotation axis, directly above and below the
black hole (less than $10^{-5}M_{\odot}$ are contained within
approximatlely $10^{\circ}$ of the rotation axis). The gravitational
radiation signal reflects the nature of the encounter, with the
amplitude of the waveforms dropping practically to zero soon after the
star is tidally disrupted. Some mass ($M_{ejected}\simeq
10^{-2}M_{\odot}$ at most, see Table~\ref{disks}), found in the outer
parts of the tidal tail formed during the initial episode of mass
transfer, has enough mechanical energy to be dynamically ejected from
the system during coalescence. We find that the amount of ejected mass
is sensitive to the value of the adiabatic index, with a sharp drop
(by more than two orders of magnitude) ocurring as it decreases below
$\Gamma=2$.

In paper~II we showed the results of dynamical calculations of
coalescence that used tidally locked binaries with an adiabatic index
$\Gamma=5/3$. Thus the effect of using an irrotational initial
condition can be gauged by directly comparing those results with the
present ones. The runs shown in paper~II also included the effects of
gravitational radiation reaction in the quadrupole approximation for
point masses, applying it to the whole star, whereas we have now done
it by identifying the self--bound core of the neutron star.
Qualitatively the outcome of the coalescence is the same for
irrotational and tidally locked systems, but there are quantitative
differences. These arise because the encounter in the case of a
tidally locked binary is more gentle, with the separation decreasing
at a slower rate once hydrodynamical effects become important. The
details can be seen by comparing the results given for run D in
paper~II (tidally locked, $\Gamma=5/3$, initial separation
$r_{i}=3.60$ and initial mass ratio $q=0.31$) with those for run D31
shown here. In the former run, the initial peak accretion rate is
lower ($\dot{M}_{max}=0.045$), the final disc mass is higher
($M_{disc}=0.226$), and the Kerr parameter of the black hole is
slightly lower ($a=0.222$) than for run D31 (where
$\dot{M}_{max}=0.074$, $M_{disc}=0.141$ and $a=0.242$), all consistent
with a less violent encounter after Roche lobe overflow.  The
accretion disc itself is not only more massive, but is located at a
larger radius, due to the higher value of total angular momentum
available in synchronized systems. This can be seen by locating the
maximum in the density (which is at $r\simeq 8$ for the tidally locked
case, see Figure~10b in paper~II, and at $r\simeq 4$ for run D31), and
the point at which the distribution of specific angular momentum $j$
flattens, marking the outer edge of the disc (at $r\simeq 13$ for run
D in paper~II, and at $r\simeq 10$ for run D31). This also makes the
maximum densities in the disc greater by at least a factor of five in
the irrotational case. The tidal tail is almost nonexistent as a
large--scale coherent structure for run D31, but can be seen clearly
in the synchronized case (see Figure~11a in paper~II). This makes for
a lower amount of dynamically ejected mass in the irrotational case
(by a factor of 200), and is due to the lower total angular momentum
contained in the system, as mentioned above. The exact factor in this
case remains somewhat uncertain, since it is sensitive to the
implementation of gravitational radiation reaction, which is slightly
different in paper~II and this work, as mentioned above. In paper~II
we quoted the dynamically ejected mass as that which had a positive
{\em total} energy, including the internal energy $u$. Analysis of
those simulations reveals that they did contain both type~I (cold,
dynamically ejected gas) and type~II ejected matter (see
section~\ref{ejected}). However, as mentioned above, there was much
more dynamically ejected matter than in run D31, and type~II matter
amounted only to $\approx$ 5~per~cent of the total. Finally, the
gravitational radiation signal is affected for the same reasons, with
the irrotational case producing a higher maximum amplitude, peak
luminosity, and total radiated energy (the differences are of 2, 14
and 7 per cent respectively).

The present results and those given in paper~III allow us to observe
general trends for all monitored quantities in irrotational systems,
as the adiabatic index is lowered from $\Gamma=3$ to $\Gamma=5/3$ (see
Tables~2 and 3 in paper~III and in this work). They can be summarized
as follows. As the compressibility increases, the peak accretion rate
increases, the disc mass drops (here we exclude the results for
$\Gamma=3$ since that case did not always imply the complete
disruption of the neutron star), the black hole has greater spin, the
peak amplitude, luminosity and efficiency of gravitational wave
emission increase, the disc lifetime decreases, and the minimum
separation attained by the binary before tidal disruption is
smaller. This last fact implies that the maximum frequency emitted by
the system in gravitational waves is higher at lower $\Gamma$, and can
be seen in Figure~\ref{gwgamma}, where we show the energy spectrum of
the gravitational wave signal for different values of $\Gamma$ at a
fixed mass ratio $q=0.31$. When the binary separation is large
compared with the stellar radius, the spectrum is close to that for a
point--mass binary, with $dE/df\propto f^{-1/3}$. When the system
becomes dynamically unstable, either through tidal effects (for low
compressibility) or because of runaway mass transfer (for high
compressibility), the power drops abruptly. This occurs at a
characteristic frequency $f_{dyn}$, which increases from $\simeq
700$~Hz to 1~kHz as the adiabatic index decreases from $\Gamma=3$ to
$\Gamma=5/3$.

Each of these consequences can be traced to the degree of central
condensation of the neutron star, to its mass--radius relationship and
hence to the way it responds to mass loss upon overflowing its Roche
lobe. The magnitude of the changes in the variables mentioned above is
not more than a factor of two. The one variable that is greatly
affected, especially at low values of $\Gamma$, is the total amount of
ejected mass (see section~\ref{ejected}). As mentioned in the
introduction, the mass ejected from this type of system might be a
source of heavy elements, if the r--process occurs, and could
contribute significantly to the observed galactic abundances. Our
numerical treatment of the coalescence does not allow us to explore
nuclear reactions, but merely estimate how much matter might leave the
system. We refer the reader to the work of Rosswog et
al. \shortcite{rosswog99,rosswog00} and Freiburghaus et
al. \shortcite{frei99b} for a detailed thermodynamical and nuclear
network calculation, in the case of double neutron star binares. The
main point in this respect in our calculations is that (i) ejection is
greatly suppressed, and practically eliminated, if the equation of
state is very soft and (ii) irrotational systems eject less mass than
tidally locked ones, by about one order of magnitude.

\begin{figure}
\psfig{width=7.5cm,file=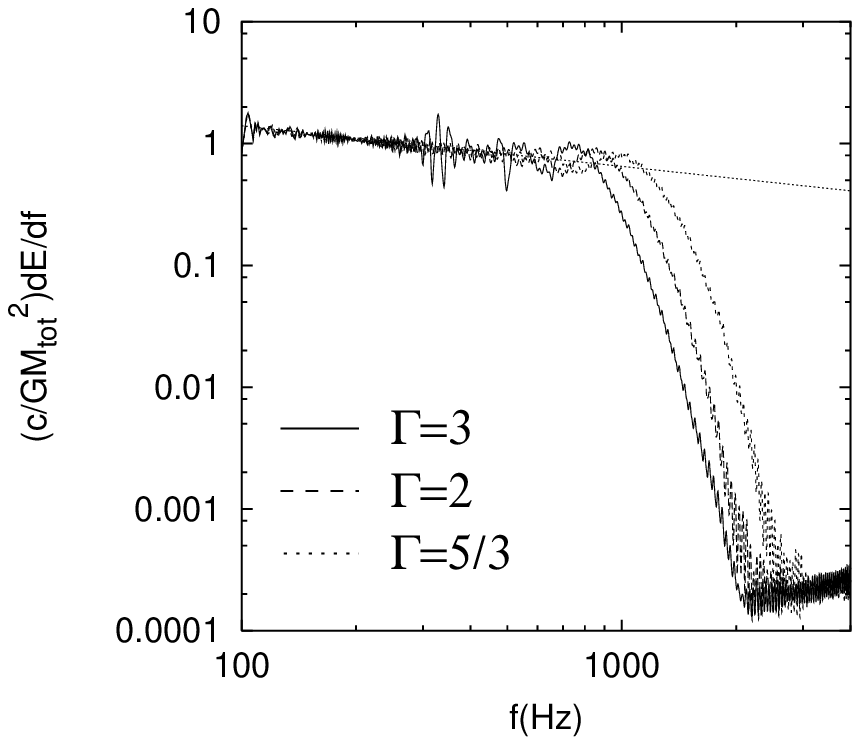,angle=0,clip=}
\caption{Gravitational waves energy spectrum $dE/df$ for dynamical
runs using irrotational binaries with initial mass ratio $q=0.31$ and
$\Gamma=3$ (solid line, run A31 from paper~III), $\Gamma=2$ (dashed
line, run C31) and $\Gamma=5/3$ (dotted line, run D31). The downward
sloping line is the result for a point--mass binary with the same mass
ratio, with $dE/df\propto f^{-1/3}$. The increased power at $f\approx
300$~Hz for $\Gamma=3$ corresponds to the return to low frequencies
after the initial mass transfer episode and the survival of the binary
(see paper~III).}
\label{gwgamma}
\end{figure}

The use of accurate equilibrium initial conditions is important in
dynamical simulations, since an initial perturbation at the start of
the calculation can propagate and affect the evolution of the
system. Using spherical neutron stars for one of our chosen mass
ratios, $q=0.31$, we have explored this effect for irrotational
binaries, for values of $\Gamma$ ranging from 3 to 5/3. We find that
the qualitative aspect of the coalescence is unaffected, but that
quantitative changes occur, all due to the instantaneous appearance of
a tidal bulge on the surface of the neutron star as the simulation
begins. The effect of this perturbation is largest at low
compressibility, since that is when a larger portion of the stellar
mass is close to the surface, and tidal effects are more
pronounced. As one decreases the value of $\Gamma$, the differences
between runs initiated with spheres (runs A31S and B31S in paper~III
and runs C31S and D31S in this work) and those that used tri--axial
ellipsoids (A31, B31, C31 and D31) become less important.

We have used a polytropic equation of state for our study in order to
use the compressibility as a free parameter. Clearly it is an
oversimplification as far as thermodynamic details are concerned, but
it allows one to explore how the system responds globally to this
variable. As we have seen, the emission of gravitational waves and the
amount of ejected mass are the two aspects that are most affected by
varying $\Gamma$. One can make the adiabatic index a function of the
density, and in this way try to model the neutron star in a more
realistic manner. This approach has been carried out by Rosswog et
al. \shortcite{rosswog99,rosswog00}, although they mainly used the
equation of state of Lattimer \& Swesty \shortcite{LS} for their
calculations. We have performed tests using this same approximation,
and have found that it is the value of $\Gamma$ at high densities that
determines the overall evolution of the system (as Rosswog et
al. did), thus fixing for example the qualitative features of the
gravitational wave emission and the amount of ejected
mass. Realistically, it would appear that the equation of state for
neutron star matter is such that the radius is nearly independent of
the mass \cite{prakash01}, and so adopting a polytropic equation of
state would require using $\Gamma=2$.

In a majority of the dynamical simulations we have performed, we have
found that massive accretion discs form, with a few tenths of a solar
mass. In all cases when this occurs, the specific angular momentum can
be approximated by a power law, with $j\propto r^{p}$. Regardless of
the value of the adiabatic index, the initial mass ratio, or the
initial distribution of angular momentum (tidally locked
vs. irrotational), we find $p\simeq 0.4-0.45$. Thus the discs are
sub--Keplerian, and are far from having a constant distribution of
specific angular momentum. This is crucial in the context of gamma-ray
bursts (see below), because it has been shown that accretion discs
around black holes can suffer from a runaway instability that destroys
them on a dynamical time--scale \cite{abram83}.  Studies over the past
two decades have shown that a number of effects can either suppress or
enhance it. Among these are (i) the spin of the black hole, (ii) the
rotation law in the disc, specified as $j\propto r^{p}$, (iii) the
effects of general relativity and (iv) the self--gravity of the
disc. Factors (i--high spin) and (ii--increasing the value of $p$)
tend to suppress the instability
\cite{wilson84,daigne,abram98,masuda98,lu00}, while (iii) and (iv)
tend to enhance it \cite{nishida96a,nishida96b,masuda98}. We note here
that all of these studies assume a softer equation of state than the
ones we have used (either using a polytrope with $\Gamma=4/3$ or a
realistic equation of state for neutron tori). Since our simulations
show that the Kerr parameter of the black hole is significant, and
that the power--law index of the distribution of specific angular
momentum is high, it would appear that these discs would not suffer
from the aforementioned instability, and would thus evolve due to
angular momentum transport on the much longer viscous (rather than
dynamical) time--scale. However, our simulations are purely Newtonian,
and thus it is impossible to include the de--stabilizing effects of
general relativity. The mass of the discs we find is apparently not
too high \cite{masuda97}, regarding the criterion for self--gravity
(the mass ratio $q_{disc}=M_{disc}/M_{BH}$ between the disc and the
black hole at the end of the calculations for irrotational binaries
ranges between 0.02, for run D20, and 0.09, for run B50 in paper~III).

The accretion discs always have a baryon--free region along the
rotation axis, above and below the black hole. This region is clear of
matter to a degree (less than $10^{-5}M_{\odot}$ within approximately
$10^{\circ}$) that would not hinder the production of a relativistic
fireball \cite{rees92,rees93}, thus powering a cosmological gamma ray
burst. The binding energy of the tori is $\simeq 10^{52}$~erg (see
e.g. Figure~\ref{energies}), and the Kerr parameter of the black hole
is $a\simeq 0.3$ at the end of the calculations, so the energy for the
burst could come either from neutrino emission from the disc or from
the spin of the black hole via the Blandford \& Znajek
\shortcite{blandford77} mechanism if the magnetic field in the torus
is strong enough and threads the black hole. The maximum extractable
energy in this latter case would be $\simeq \epsilon_{BZ}10^{53}$~erg,
where $\epsilon_{BZ}$ is the MHD efficiency factor. In either case,
one would expect the disc to survive for a time--scale comparable to
the duration of the burst, i.e.  on the order of seconds. This is why
the previously mentioned result concerning the power law distribution
of angular momentum and the accompanying dynamical stability of the
disc is so important. The short time--scales and rapid variability
involved in a small ($\approx 100$~km) accretion disc around a black
hole make these systems attractive candidates for the central engines
of short gamma ray bursts, as we found in our preliminary studies
\cite{kl98}, and have now confirmed in the present series of papers
for a wide variety of initial conditions, varying the stiffness of the
equation of state, the initial mass ratio in the binary and the
distribution of angular momentum in the system.

We note that the mounting observational evidence in favor of massive
stars being GRB progenitors \cite{galama01} does not exclude compact
mergers as sources, simply because all observed afterglows, from which
the inferences about the environment where the bursts occur come,
correspond to long bursts. If compact mergers do in fact produce GRBs,
spectacular confirmation about the nature of the source could be
obtained through the detection of a coincident gravitational wave
signal, even if the final coalescence waveform is outside the
frequency band of detectors such as LIGO. One could observe the final
minutes of the inspiral phase as the orbital frequency increases,
leaves the LIGO band, and then search for a coincident GRB.

\section*{Acknowledgments}

This work has benefited greatly from conversations with W\l odzimierz
Klu\'{z}niak, Frederic Rasio, Maximilian Ruffert and Lars Bildsten. I
thank the referee for a thorough reading of the manuscript and a
prompt report, and for pointing out the error in the calculation of
the Kerr parameter of the black hole. Support for this work was
provided by CONACyT (27987E) and DGAPA--UNAM (IN-119998).

\appendix

\section{Computation of the Kerr parameter of the black hole}

The Kerr parameter $a$ of the black hole was incorrectly calculated
for the results presented in papers~II and III. In this appendix we
show explicitly the correct derivation of $J_{\rm BH}^{spin}$, and the
corrected values for all the runs performed in papers~II and III.

When a gas (SPH) particle crosses the accretion boundary of the black
hole, set at the Schwarzschild radius $r_{Sch}=2GM_{\rm BH}/c^{2}$, we
update the mass and velocity of the black hole so as to ensure
conservation of mass and total linear momentum, i.e.

\begin{equation}
M_{\rm BH}^{\prime} = M_{\rm BH} + m_{i},
\end{equation}
and
\begin{equation}
M_{\rm BH}^{\prime} \vec{v}_{\rm BH}^{\prime}= M_{\rm BH} \vec{v}_{\rm
BH} + m_{i} \vec{v}_{i},
\end{equation}
where primed quantities refer to values after the particle has been
accreted and removed from the simulation.

The conservation of total angular momentum reads:

\begin{equation}
\vec{r}_{i} \times m_{i} \vec{v}_{i} + \vec{r}_{\rm BH} \times M_{\rm
BH} \vec{v}_{\rm BH}=\vec{r}_{\rm BH}^{\prime} \times M_{\rm
BH}^{\prime} \vec{v}_{\rm BH}^{\prime} + \vec{J}_{\rm BH}^{spin},
\end{equation}
where $J_{\rm BH}^{spin}$ is the spin angular momentum gained by the
black hole because of the accretion. A fraction of the particle's
angular momentum contributes to the orbital angular momentum of the
black hole, and the rest to its spin. In practice, we found that the
latter term dominates, and that our error was due mainly to not taking
into account the angular momentum lost to gravitational waves (which
is most important in the early stages of the simulation, before the
neutron star has been disrupted). Table~\ref{correctj} shows the
correct values for the Kerr parameter of the black hole for the runs
shown in paper~II (A through E) and for those presented in paper~III
(A50, A31, A31S, A20, B50, B31S, B31 and B20).

\begin{table}
 \caption{Kerr parameter of the black hole at the end of the dynamical
 simulations for the runs presented originally in papers II and III}
 \label{correctj}
 \begin{tabular}{@{}lcccc}
  Run & $\Gamma$ & $q$ & $J_{\rm BH}c/G M_{\rm BH}^{2}$ & Reference \\
  A     & 5/3 & 1.00 & 0.448 & paper~II \\
  B     & 5/3 & 0.80 & 0.409 & paper~II \\
  C     & 5/3 & 0.31 & 0.232 & paper~II \\
  D     & 5/3 & 0.31 & 0.222 & paper~II \\
  E     & 5/3 & 0.10 & 0.097 & paper~II \\
  A50   & 3.0 & 0.50 & 0.339 & paper~III \\
  A31   & 3.0 & 0.31 & 0.226 & paper~III \\
  A31S  & 3.0 & 0.31 & 0.238 & paper~III \\
  A20   & 3.0 & 0.20 & 0.156 & paper~III \\
  B50   & 2.5 & 0.50 & 0.339 & paper~III \\
  B31   & 2.5 & 0.31 & 0.244 & paper~III \\
  B31S  & 2.5 & 0.31 & 0.247 & paper~III \\
  B20   & 2.5 & 0.20 & 0.167 & paper~III \\

 \end{tabular}

\end{table}

\label{lastpage}


\begin{thebibliography}{}

   \bibitem[\protect\citename{Abramovici et al.\ }1992]{abram92}
   Abramovici M. et al. 1992, Science, 256, 325

   \bibitem[\protect\citename{Abramowicz, Calvani \& Nobili
   }1983]{abram83} Abramovicz M.~A., Calvani M., Nobili L. 1983, Nat.,
   302, 597

   \bibitem[\protect\citename{Abramowicz, Karas \& Lanza
   }1998]{abram98} Abramovicz M.~A., Karas V., Lanza A. 1998, A\& A,
   331, 1143

   \bibitem[\protect\citename{Ayal et al.\ }2001]{ayal01} Ayal S.,
   Piran T., Oechslin R., Davies M.~B., Rosswog S. 2001, ApJ, 550, 846

   \bibitem[\protect\citename{Balsara }1995]{balsara} Balsara D. 1995,
   J. Comp. Phys., 121, 357

   \bibitem[\protect\citename{Baumgarte et al.\ }1997]{baumgarte97}
   Baumgarte T.~W., Cook G.~B., Scheel M.~A., Shapiro S.~L., Teukolsky
   S.~A. 1997, Phys. Rev. Lett., 79, 1182

   \bibitem[\protect\citename{Belczy\'{n}ski \& Bulik
   }1999]{belczynski} Belczy\'{n}ski K., Bulik T. 1999, A\&A, 346, 91

   \bibitem[\protect\citename{Bethe \& Brown }1998]{bethe} Bethe
   H.~A., Brown G.~E. 1998, ApJ, 506, 780

   \bibitem[\protect\citename{Bildsten \& Cutler }1992]{bildsten}
   Bildsten L., Cutler C. 1992, ApJ, 400, 175

   \bibitem[\protect\citename{Blanchet et al.\ }1995]{blanchet95}
   Blanchet L., Damour T., Iyer B.~R., Will C.~M., Wiseman A.~G. 1995,
   Phys. Rev. Lett., 74, 3515

   \bibitem[\protect\citename{Blandford \& Znajek }1977]{blandford77}
   Blandford R.~D., Znajek R.~L. 1977, MNRAS, 179, 433

   \bibitem[\protect\citename{Bradaschia et al.\ }1990]{brad}
   Bradaschia C. et al. 1990, Nucl. Instrum. Methods Phys. Res.,
   Sect. A, 289, 518

   \bibitem[\protect\citename{Chandrasekhar }1969]{ch69} Chandrasekhar
   S. 1969, Ellipsoidal figures of equilibrium, (New Haven: Yale
   Univ. Press)

   \bibitem[\protect\citename{Cutler et al.\ }1994]{cutler} Cutler
   C. et al. 1993, Phys. Rev. Lett., 70, 2984

   \bibitem[\protect\citename{Daigne \& Mochkovitch }1997]{daigne}
   Daigne F., Mochkovitch R. 1997, MNRAS, 285, L15

   \bibitem[\protect\citename{Davies et al.\ }1994]{davies} Davies
   M.~B., Benz W., Piran T., Thielemann F.--K. 1994, ApJ, 431, 742

   \bibitem[\protect\citename{Djorgovski et al.\ }1998]{djorgovski98}
   Djorgovski S.~G., Kulkarni S.~R., Bloom J.~S., Goodrich R., Frail
   D.~A., Piro L., Palazzi E. 1998, ApJ, 508, L17

   \bibitem[\protect\citename{Eichler et al.\ }1989]{eichler} Eichler
   D., Livio M., Piran T., Schramm D.~N. 1989, Nat., 340, 126

   \bibitem[\protect\citename{Faber \& Rasio }2000]{faber00} Faber
   J.~A., Rasio F.~A. 2000, Phys. Rev. D, 62, 064012

   \bibitem[\protect\citename{Faber, Rasio \& Manor }2000]{faber01}
   Faber J.~A., Rasio F.~A., Manor J.~B. 2001, Phys. Rev. D, 63,
   044012

   \bibitem[\protect\citename{Finn }1989]{finn} Finn L.S. 1989 in
   Evans C.~R., Finn L.~S., Hobill D.~W., eds, Frontiers of Numerical
   Relativity, Cambridge Univ. Press. Cambridge, p.126

   \bibitem[\protect\citename{Fishman \& Meegan }1995]{fishman95}
   Fishman G.~J., Meegan C.~A. 1995, ARA\&A, 33, 415

   \bibitem[\protect\citename{Frail et al.\ }2001]{frail01} Frail
   D.~A. et al. 2001, Nat., submitted, astro-ph/0102282

   \bibitem[\protect\citename{Freiburghaus et al.\ }1999a]{frei99a}
   Freiburghaus C., Rembges J.--F., Rauscher T., Kolbe E., Thielemann
   F.--K. 1999a, ApJ, 516, 381

   \bibitem[\protect\citename{Freiburghaus et al.\ }1999b]{frei99b}
   Freiburghaus C., Rosswog S., Thielemann F.--K. 1999b, ApJ, 525,
   L121

   \bibitem[\protect\citename{Fryer, Woosley \& Hartmann
   }1999a]{fryer99a} Fryer C.~L., Woosley W.~E., Hartmann D.~H.,
   1999a, ApJ, 526, 152

   \bibitem[\protect\citename{Fryer et al.\ }1999b]{fryer99b} Fryer
   C.~L., Woosley W.~E., Herant M., Davies M.~B. 1999b, ApJ, 520, 650


   \bibitem[\protect\citename{Galama \& Wijers }2001]{galama01} Galama
   T. J., Wijers R. A. M. J. 2001, ApJ, 549, L209

   \bibitem[\protect\citename{Goodman }1986]{goodman86} Goodman
   J. 1986, ApJ, 308, L46

   \bibitem[\protect\citename{Goodman }1987]{goodman87} Goodman J.,
   Dar A., Nussinov S. 1987, ApJ, 314, L7

   \bibitem[\protect\citename{Gourgoulhon et al.\
   }2001]{gourgoulhon01} Gourgoulhon E., Grandclement P., Taniguchi
   K., Marck J.--A., Bonazzola S. 2001, Phys. Rev. D, 63, 064029

   \bibitem[\protect\citename{Harrison et al.\ }1999]{harrison99}
   Harrison F.~A. et al. 1999, ApJ, 523, L121

   \bibitem[\protect\citename{Hulse \& Taylor }1975]{hulse75} Hulse
   R.~A., Taylor J.~H. 1975, ApJ, 195, L51

   \bibitem[\protect\citename{Janka et al.\ }1999]{janka99} Janka
   H.--Th., Eberl T., Ruffert M., Fryer C.~L. 1999, ApJ, 527, L39

   \bibitem[\protect\citename{Jarozy\'{n}ski }1993]{jaro93}
   Jaroszy\'{n}ski M. 1993, Acta Astron., 43, 183

   \bibitem[\protect\citename{Jarozy\'{n}ski }1996]{jaro96}
   Jaroszy\'{n}ski M. 1996, A\&A, 305, 839

   \bibitem[\protect\citename{Kalogera et al.\ }2001]{kalogera01a}
   Kalogera V., Narayan R., Spergel D.~N., Taylor J.~H. 2001, ApJ,
   556, 340

   \bibitem[\protect\citename{Kalogera \& Belczy\'{n}ski
   }2001]{kalogera01b} Kalogera V., Belcy\'{n}ski K. 2001, in
   Centrella J., ed., AIP Proc. 575, Astrophysical sources for
   ground--based gravitational wave detectors, AIP, New York, p. 107,
   astro-ph/0101047

   \bibitem[\protect\citename{Kidder, Will \& Wiseman }1992]{kidder92}
   Kidder L.~E., Will C.~M., Wiseman A.~G., 1992, Class. Quantum
   Grav., 9, L125

   \bibitem[\protect\citename{Klu\'{z}niak \& Lee }1998]{kl98}
   Klu\'{z}niak W., Lee W.~H. 1998, ApJ, 454, L53

   \bibitem[\protect\citename{Klu\'{z}niak \& Ruderman }1998]{kr98}
   Klu\'{z}niak W., Ruderman M. 1998, ApJ, 505, L113

   \bibitem[\protect\citename{Kochanek }1992]{kochanek} Kochanek
   C. 1992, ApJ, 398, 234

   \bibitem[\protect\citename{Kouveliotou et al.\ }1995]{kouveliotou}
   Kouveliotou C., Koshut T., Briggs M.~S., Pendleton G.~N., Meegan
   C.~A., Fishman G.~J., Lestrade J.~P. 1995, in Kouveliotou C.,
   Briggs M.~F., Fishman G.~J., eds., AIP Proc. 384, Gamma Ray Bursts,
   AIP, New York, p.~42

   \bibitem[\protect\citename{Kulkarni et al.\ }1998]{kulkarni98}
   Kulkarni S.~R. et al. 1998, Nat., 393, 35

   \bibitem[\protect\citename{Kulkarni et al.\ }1999]{kulkarni99}
   Kulkarni S.~R. et al. 1999, Nat., 398, 389

   \bibitem[\protect\citename{Lai, Rasio \& Shapiro }1993a]{LRSa} Lai
   D., Rasio F.~A., Shapiro S.~L. 1993a, ApJ, 406, L63 (LRSa)

   \bibitem[\protect\citename{Lai, Rasio \& Shapiro }1993b, hereafter
   LRSb]{LRSb} Lai D., Rasio F.~A., Shapiro S.~L. 1993b, ApJS, 88, 205
   (LRSb)

   \bibitem[\protect\citename{Lai, Rasio \& Shapiro }1994]{LRS94} Lai
   D., Rasio F.~A., Shapiro S.~L. 1994, ApJ, 437, 742

   \bibitem[\protect\citename{Landau \& Lifshitz }1975]{landau75}
   Landau L.D., Lifshitz E.M. 1975, The Classical Theory of Fields,
   Heinemann, Oxford

   \bibitem[\protect\citename{Lattimer \& Schramm }1974]{LS74}
   Lattimer J.~M., Schramm D.~N. 1974, ApJ, 192, L145

   \bibitem[\protect\citename{Lattimer \& Schramm }1976]{LS76}
   Lattimer J.~M., Schramm D.~N. 1976, ApJ, 210, 549

   \bibitem[\protect\citename{Lattimer \& Swesty }1991]{LS} Lattimer
   J.~M., Swesty D. 1991, Nuc. Phys. A, 535, 331

   \bibitem[\protect\citename{Lee, Wijers \& Brown }2000]{lee00} Lee
   H.~K., Wijers R.~A.~M.~J., Brown G.~E. 2000, Phys. Rep. 325, 83

   \bibitem[\protect\citename{Lee }1998]{phd} Lee W.~H. 1998, PhD
   Thesis, University of Wisconsin

   \bibitem[\protect\citename{Lee \& Klu\'{z}niak }1995]{acta} Lee
   W.~H., Klu\'{z}niak W. 1995, Acta Astron., 45, 705

   \bibitem[\protect\citename{Lee \& Klu\'{z}niak }1997]{hunt4} Lee
   W.~H., Klu\'{z}niak W. 1997, in Meegan C., Preece R., Koshut
   P. eds., AIP Proc. 428, Gamma Ray Bursts, AIP, New York, p.~798

   \bibitem[\protect\citename{Lee \& Klu\'{z}niak }1999a]{LKI} Lee
   W.~H., Klu\'{z}niak W. 1999a, ApJ, 526, 178 (paper~I)

   \bibitem[\protect\citename{Lee \& Klu\'{z}niak }1999b, hereafter
   papers I \& II]{LKII} Lee W.~H., Klu\'{z}niak W. 1999b, MNRAS, 308,
   780 (paper~II)

   \bibitem[\protect\citename{Lee }2000, hereafter paper III]{LKIII}
   Lee W.~H. 2000, MNRAS, 318, 606 (paper~III)

   \bibitem[\protect\citename{Lipunov, Postnov \& Prokhorov
   }1997]{lipunov} Lipunov V.~M., Postnov K.~A., Prokhorov M.~E.,
   1997, New Ast., v.2, 43

   \bibitem[\protect\citename{Lombardi, Rasio \& Shapiro
   }1997]{lombardi97} Lombardi J.~C., Rasio F.~A., Shapiro S.~L.,
   1997, Phys. Rev. D, 56, 3416

   \bibitem[\protect\citename{Lu et al.\ }2000]{lu00} Lu Y., Cheng
   K.~S., Yang L.~T., Zhang L. 2000, MNRAS, 314, 453

   \bibitem[\protect\citename{MacFadyen \& Woosley }1999]{macfadyen}
   MacFadyen A., Woosley S.~E. 1999, ApJ, 524, 262

   \bibitem[\protect\citename{Masuda \& Eriguchi }1997]{masuda97}
   Masuda N., Eriguchi Y. 1997, ApJ, 489, 804

   \bibitem[\protect\citename{Masuda, Nishida \& Eriguchi
   }1998]{masuda98} Masuda N., Nishida S., Eriguchi Y. 1998, MNRAS,
   297, 1139

   \bibitem[\protect\citename{Meegan et al.\ }1992]{meegan} Meegan
   C.~A., Fishman G.J., Wilson R.~B., Horack J.~M., Brock M.~N.,
   Paciesas W.~S., Pendleton G.~N., Kouveliotou, C. 1992, Nat., 355,
   143

   \bibitem[\protect\citename{Metzger et al.\ }1997]{metzger97}
   Metzger M.~R., Djorgovski S.~G., Kulkarni S.~R., Steidel C.~C.,
   Adelberger K.~L., Frail D.~A., Costa E., Frontera F. 1997, Nat.,
   387, 878

   \bibitem[\protect\citename{M\'{e}sz\'{a}ros \& Rees }1992]{rees92}
   M\'{e}sz\'{a}ros P., Rees M.~J. 1992, MNRAS, 257, 29P

   \bibitem[\protect\citename{M\'{e}sz\'{a}ros \& Rees }1993]{rees93}
   M\'{e}sz\'{a}ros P., Rees M.~J. 1993, ApJ, 405, 278

   \bibitem[\protect\citename{M\'{e}sz\'{a}ros \& Rees
   }1997a]{rees97a} M\'{e}sz\'{a}ros P., Rees M.~J. 1997a, ApJ, 476,
   232

   \bibitem[\protect\citename{M\'{e}sz\'{a}ros \& Rees
   }1997b]{rees97b} M\'{e}sz\'{a}ros P., Rees M.~J. 1997b, ApJ, 482,
   L29

   \bibitem[\protect\citename{M\'{e}sz\'{a}ros, Rees \& Wijers
   }1999]{mesz99} M\'{e}sz\'{a}ros P., Rees M.~J., Wijers
   R.~A.~M.~J. 1999, New Ast., 4, 303

   \bibitem[\protect\citename{Meyer \& Brown }1997]{meyer97} Meyer
   B.~S., Brown J.~S. 1997, ApJS, 112, 199

   \bibitem[\protect\citename{Mochkovitch et al.\ }1993]{moch93}
   Mochkovitch R., Hernanz M., Isern J., Martin X. 1993, Nat., 361,
   236

   \bibitem[\protect\citename{Mochkovitch et al.\ }1995]{moch95}
   Mochkovitch R., Hernanz M., Isern J., Loiseau S. 1995, A\& A,293,
   803

   \bibitem[\protect\citename{Monaghan }1992]{monaghan92} Monaghan
   J.~J. 1992, ARA\&A, 30, 543

   \bibitem[\protect\citename{Nakamura \& Oohara }1989]{nakamura89}
   Nakamura T., Oohara K. 1989, Prog. Theor. Phys., 82, 1066

   \bibitem[\protect\citename{Nakamura \& Oohara }1991]{nakamura91}
   Nakamura T., Oohara K. 1991, Prog. Theor. Phys., 86, 73

   \bibitem[\protect\citename{Narayan, Paczy\'{n}ski \& Piran
   }1992]{narayan92} Narayan R., Paczy\'{n}ski B., Piran T., 1992,
   ApJ, 395, L83

   \bibitem[\protect\citename{Narayan, Piran \& Shemi
   }1991]{narayan91} Narayan R., Piran T., Shemi A. 1991, ApJ, 379,
   L17

   \bibitem[\protect\citename{Nishida et al.\ }1996]{nishida96a}
   Nishida S., Lanza A., Eriguchi Y., Abramowicz M.~A. 1996, MNRAS,
   278, L41

   \bibitem[\protect\citename{Nishida \& Eriguchi }1996]{nishida96b}
   Nishida S., Eriguchi Y. 1996, ApJ, 461, 320

   \bibitem[\protect\citename{Oohara \& Nakamura }1989]{oohara89}
   Oohara K., Nakamura T. 1989, Prog. Theor. Phys., 82, 535

   \bibitem[\protect\citename{Oohara \& Nakamura }1990]{oohara90}
   Oohara K., Nakamura T. 1990, Prog. Theor. Phys., 83, 906

   \bibitem[\protect\citename{Oohara \& Nakamura }1992]{oohara92}
   Oohara K., Nakamura T. 1992, Prog. Theor. Phys., 88, 307

   \bibitem[\protect\citename{Oohara \& Nakamura }1999]{oohara99}
   Oohara K., Nakamura T. 1999, Prog. Theor. Phys. Supp., 136, 270

   \bibitem[\protect\citename{Paczy\'{n}ski }1986]{paczynski86}
   Paczy\'{n}ski B. 1986, ApJ, 308, L43

   \bibitem[\protect\citename{Paczy\'{n}ski }1991]{paczynski91}
   Paczy\'{n}ski B. 1991, Acta Astron., 41, 257

   \bibitem[\protect\citename{Popham, Woosley \& Fryer }1999]{popham}
     Popham R., Woosley S.~E, Fryer C. 1999, ApJ, 518, 356

   \bibitem[\protect\citename{Portegies Zwart \& Yungelson
   }1998]{portyun} Portegies Zwart S.~F., Yungelson L.~F. 1998, A\&A,
   372, 173

   \bibitem[\protect\citename{Prakash \& Lattimer }2001]{prakash01}
   Prakash M., Lattimer J.~M. 2001, ApJ, 550, 426

   \bibitem[\protect\citename{Rasio \& Shapiro }1992]{RS92} Rasio
   F.~A., Shapiro S.~L. 1992, ApJ, 401, 226 (RS92)

   \bibitem[\protect\citename{Rasio \& Shapiro }1994]{RS94} Rasio
   F.~A., Shapiro S.~L. 1994, ApJ, 432, 242 (RS94)

   \bibitem[\protect\citename{Rasio \& Shapiro }1995, hereafter RS92,
   RS94, RS95 respectively]{RS95} Rasio F.~A., Shapiro S.~L. 1995,
   ApJ, 438, 887 (RS95)

   \bibitem[\protect\citename{Rees \& M\'{e}sz\'{a}ros }1992]{rm92}
   Rees M.~J., M\'{e}sz\'{a}ros P. 1992, MNRAS, 258, 41P

   \bibitem[\protect\citename{Rosswog et al.\ }1999]{rosswog99}
   Rosswog S., Liebend\"{o}rfer M., Thielemann F.--K., Davies M.~B.,
   Benz W., Piran T. 1999, A\&A, 341, 499

   \bibitem[\protect\citename{Rosswog et al.\ }2000]{rosswog00}
   Rosswog S., Davies M.B., Thielemann F.--K., Piran T. 2000, A\&A,
   360, 171

   \bibitem[\protect\citename{Ruderman, Tao \& Klu\'{z}niak
   }2000]{rtk00} Ruderman M., Tao L., Klu\'{z}niak W. 2000, ApJ, 542,
   243

   \bibitem[\protect\citename{Ruffert, Janka \& Sch\"{a}fer
   }1996]{ruffert96} Ruffert M., Janka H.--Th., Sch\"{a}fer G. 1996,
   A\&A, 311, 532

   \bibitem[\protect\citename{Ruffert et al.\ }1997]{ruffert97}
   Ruffert M., Janka H.--Th., Takahashi K., Sch\"{a}fer G. 1997, A\&A,
   319, 122

   \bibitem[\protect\citename{Ruffert \& Janka }1999]{ruffert99}
   Ruffert M., Janka H.--Th. 1999, A\&A, 344, 573

   \bibitem[\protect\citename{Salmonson, Wilson \& Mathews
   }2001]{salmonson01} Salmonson J.~D., Wilson J.~R., Mathews
   G.~J. 2001, ApJ, 553, 471

   \bibitem[\protect\citename{Shibata }1999]{shibata99} Shibata
   M. 1999, Phys. Rev. D, 60, 104052

   \bibitem[\protect\citename{Shibata \& Ury\={u} }2000]{shibata00}
   Shibata M., Ury\={u} K. 2000, Phys. Rev. D, 61, 064001

   \bibitem[\protect\citename{Spruit }1999]{spruit99} Spruit H. 1999,
   A\&A, 341, L1

   \bibitem[\protect\citename{Stairs et al.\ }1998]{stairs98} Stairs
   I.~H., Arzoumanian Z., Camilo F., Lyne A.~G., Nice D.~J., Taylor
   J.~H., Thorsett S.~E., Wolszczan A., 1998, ApJ, 505, 352

   \bibitem[\protect\citename{Stanek et al.\ }1999]{stanek99} Stanek
   K.~Z., Garnavich P.~M., Kaluzny J., Pych W., Thompson I. 1999, ApJ,
   522, L39


   \bibitem[\protect\citename{Symbalisty \& Schramm }1989]{symb82}
   Symbalisty E.~M.~D., Schramm D.~N. 1982, Astrophyscial Letters, 22,
   143

   \bibitem[\protect\citename{Taylor et al.\ }1992]{taylor} Taylor
   J.~H., Wolszczan A., Damour T., Weisberg J.M. 1992, Nat., 355, 132

   \bibitem[\protect\citename{Thompson }1994]{thompson} Thompson
   C. 1994, MNRAS, 270, 480

   \bibitem[\protect\citename{Tutukov \& Yungelson }1993]{tutukov}
   Tutukov A.~V., Yungelson L.~R. 1993, MNRAS, 260, 675

   \bibitem[\protect\citename{Ury\={u} \& Eriguchi }1999]{uryu99}
   Ury\={u} K., Eriguchi Y. 1999, MNRAS, 303, 329

   \bibitem[\protect\citename{Ury\={u} \& Eriguchi }2000]{uryu00}
   Ury\={u} K., Eriguchi Y. 2000, Phys. Rev. D, 2000, 61, 124023

   \bibitem[\protect\citename{Usov }1992]{usov92} Usov V.~V. 1992,
   Nat., 357, 472

   \bibitem[\protect\citename{Usui, Ury\={u} \& Eriguchi
   }2000]{usui00} Usui F., Ury\={u} K., Eriguchi Y. 2000,
   Phys. Rev. D, 2000, 61, 024039

   \bibitem[\protect\citename{van Paradijs, Kouveliotou \& Wijers
   }2000]{vanpar00} van Paradijs J., Kouveliotou C., Wijers
   R.~A.~M.~J. 2000, ARA\&A, 38, 379

   \bibitem[\protect\citename{Wheeler }1971]{wheeler71} Wheeler
   J.~A. 1971, Pontificae Acad. Sci. Scripta Varia, 35, 539

   \bibitem[\protect\citename{Wilson }1984]{wilson84} Wilson
   D.~B. 1984, Nat., 312, 620

   \bibitem[\protect\citename{Wilson, Mathews \& Marronetti
   }1996]{wilson96} Wilson J.~R., Mathews G.~J., Marronetti P., 1996,
   Phys. Rev. D, 54, 1317

   \bibitem[\protect\citename{Witt et al.\ }1994]{witt} Witt H.~J.,
   Jaroszy\'{n}ski M., Haensel P., Paczy\'{n}ski B., Wambsganss
   J. 1994, ApJ, 422, 219

   \bibitem[\protect\citename{Wolszczan }1991]{wolszczan91} Wolszczan
   A. 1991, Nat., 350, 688

   \bibitem[\protect\citename{Woosley }1993]{woosley93} Woosley
   S.~E. 1993, ApJ, 405, 273

   \bibitem[\protect\citename{Zhang \& Fryer }2001]{zhang01} Zhang W.,
   Fryer C.~L. 2001, ApJ, 550, 357

   \bibitem[\protect\citename{Zhuge, Centrella \& McMillan
   }1994]{zhuge94} Zhuge X., Centrella J.~M., McMillan S.~L.~W., 1994,
   Phys. Rev. D, 50, 6247

   \bibitem[\protect\citename{Zhuge, Centrella \& McMillan
   }1996]{zhuge96} Zhuge X., Centrella J.~M., McMillan S.~L.~W., 1996,
   Phys. Rev. D, 54, 7261.


\end{thebibliography}
\end{document}